\documentclass[preprint,12pt,nopreprintline]{elsarticle}

\usepackage{graphicx}
\usepackage{amsmath}
\usepackage{amssymb}
\usepackage{float}
\usepackage{booktabs}
\usepackage{hyperref}
\hypersetup{hidelinks}
\usepackage{longtable}
\usepackage{etoolbox}
\makeatletter
\patchcmd{\pprintMaketitle}{\Large\@title\par\vskip18pt}{\Large\@title\par\vskip10pt}{}{\PackageError{layout}{Title patch failed}{}}
\patchcmd{\pprintMaketitle}{\normalsize\elsauthors\par\vskip10pt}{\normalsize\elsauthors\par\vskip6pt}{}{\PackageError{layout}{Authors patch failed}{}}
\patchcmd{\pprintMaketitle}{\footnotesize\itshape\elsaddress\par\vskip36pt}{\footnotesize\itshape\elsaddress\par\vskip12pt}{}{\PackageError{layout}{Address patch failed}{}}
\makeatother

\newcommand{\dgm}{\mathsf{dgm}}
\newcommand{\PI}{\mathrm{PI}}
\newcommand{\ad}{\mathbf{a}}
\newcommand{\e}{\mathbf{e}}

\begin{document}
\raggedbottom
\setlength{\emergencystretch}{2em}

\begin{frontmatter}

\title{From Fourier Amplitudes to Persistence Images for Impact-Echo Defect Detection}

\author[kyushu]{Keunsu Kim}
\ead{kim.keunsu.752@m.kyushu-u.ac.jp}

\author[ajou]{Suyoung Choi\corref{cor1}}
\ead{schoi@ajou.ac.kr}

\author[kyoto]{Meiyan Kang\corref{cor1}}
\ead{kang.meiyan.2c@kyoto-u.ac.jp}

\affiliation[kyushu]{organization={Institute of Mathematics for Industry, Kyushu University},
            city={Fukuoka},
            country={Japan}}

\affiliation[ajou]{organization={Department of Mathematics, Ajou University},
            city={Suwon},
            country={Republic of Korea}}

\affiliation[kyoto]{organization={Graduate School of Science, Kyoto University},
            city={Kyoto},
            country={Japan}}

\cortext[cor1]{Corresponding authors.}

\begin{abstract}
\small
Impact-echo testing is a nondestructive testing method used to detect internal defects in concrete structures. 
However, boundary reflections, structured interference, and measurement noise can produce complex Fourier amplitude spectra, making defect classification difficult. 
We constructed a two-dimensional frequency--amplitude persistence image (FA-PI) from the Fourier amplitude spectrum.
Using a fixed parameterization of exact multi-parameter persistent homology, each retained Fourier component is represented with frequency-bin location as one coordinate and amplitude as the other, and Gaussian kernels are then applied in this two-dimensional space. This construction has a direct analytical connection to the corresponding one-dimensional Gaussian-smoothed Fourier amplitude spectrum.
FA-PI was evaluated on an experimental backfill-void dataset and a public bridge-deck impact-echo dataset, and compared with raw and Gaussian-smoothed Fourier amplitude spectra using a linear classifier with a Softmax output and identical signal-level data splits.
The evaluation included original signals and signals corrupted by Gaussian noise, sinusoidal interference, or mixed noise, with matching noise types and signal-to-noise ratios during training, validation, and testing. 
Performance was assessed using accuracy and macro-averaged precision, recall, and F1-score. Across the two datasets, FA-PI consistently outperformed the raw and Gaussian-smoothed Fourier baselines in the noisy-condition summaries.
These findings indicate that representing Fourier spectral components in a two-dimensional frequency--amplitude space can improve impact-echo defect classification beyond frequency-axis smoothing alone.

\end{abstract}

\end{frontmatter}

\section{Introduction}
\label{sec:introduction}
Impact-echo testing is an established nondestructive method for evaluating internal defects in concrete structures using the transient response to mechanical impact \cite{sansalone1986impact, carino2001impact, carino2015impact}.
When a stress wave reaches an internal interface with an acoustic impedance contrast, such as a concrete--air boundary, part of the wave is reflected \cite{krautkramer2013ultrasonic}.
Conventional analysis examines the Fourier amplitude spectrum to identify characteristic frequencies associated with structural thickness and internal reflecting interfaces.
In practice, however, the measured response contains contributions from multiple wave modes, external boundaries, structural geometry, and measurement noise.
These contributions can produce overlapping or ambiguous spectral peaks, complicating the interpretation of defect-related responses.
Reliable defect assessment therefore requires signal analysis methods that can distinguish sound and defective conditions despite this spectral complexity.

Several signal-processing approaches have been developed to improve the interpretation of complex impact-echo responses.
\citet{yeh2008wavelet} combined wavelet and Fourier analysis to first narrow the frequency range associated with the echo response and then determine the peak frequency more precisely. \citet{zhang2012eemd} decomposed impact-echo responses into separate oscillatory components to isolate defect-related echoes from strong surface-wave contributions. These studies show how complementary representations and signal decomposition can reveal spectral structure more clearly.

\citet{dorafshan2020evaluation,dorafshan2020deep} evaluated deep learning models using one-dimensional impact-echo signals and two-dimensional short-time Fourier transform (STFT) spectrograms for bridge-deck evaluation with and without overlays.
\citet{sanostudy} similarly used STFT-based time--frequency representations with a convolutional neural network (CNN) for concrete defect detection.
In our previous study, we combined continuous wavelet transform (CWT) images with a CNN for tunnel-lining backfill void detection \cite{lee2024void}.
Approaches based on handcrafted features have also been investigated, with \citet{wu2025grouting} and \citet{lee2025machine} constructing input vectors from features derived from time, frequency, and time--frequency representations of impact-echo signals for machine learning.
Taken together, these studies demonstrate the role of signal representation and feature extraction in learning-based impact-echo assessment. Against this background, we constructed a two-dimensional frequency--amplitude representation for impact-echo defect classification and evaluated it against raw and Gaussian-smoothed Fourier amplitude spectra.

The Fourier amplitude spectrum explicitly associates each frequency-bin location with an amplitude, providing a direct link to conventional impact-echo spectral analysis.
To construct the proposed representation, we used exact multi-parameter persistent homology (EMPH), introduced in our earlier work \cite{kim2026exact}, which provides an analytical mapping from Fourier representations to persistence diagrams.
Persistence images provide a vector representation of such diagrams for machine learning \cite{adams2017persistence} and have also been used in EMPH-based classification \cite{kim2024filtration}.
Under the EMPH parameterization used in this study, the two persistence diagram coordinates have a direct spectral interpretation: the birth coordinate equals the retained frequency-bin index, while the persistence coordinate is proportional to the corresponding Fourier amplitude.

Based on this correspondence, we placed persistence-weighted Gaussian kernels at the birth--persistence coordinates to form a continuous two-dimensional surface. Sampling this surface on a finite grid yields the frequency--amplitude persistence image (FA-PI). Because Fourier amplitude is incorporated into both the location and weighting of the Gaussian kernels, FA-PI differs from frequency-axis smoothing alone. We established that marginalizing the continuous surface over the persistence coordinate yields the corresponding one-dimensional Gaussian-smoothed Fourier amplitude spectrum up to a known scale factor.

FA-PI was compared with raw and one-dimensional Gaussian-smoothed Fourier amplitude spectra using the same type of linear classifier with a Softmax output and identical signal-level data splits.
The evaluation was conducted separately on an experimental backfill-void dataset previously used in our CWT--CNN study \cite{lee2024void} and a public bridge-deck dataset \cite{dorafshan2020dataset}.
Unlike our earlier study, we evaluated both original signals and signals with artificially added noise.
For each noisy condition, the noise type and signal-to-noise ratio were matched across training, validation, and testing.

CWT--Softmax was included as an additional time--frequency benchmark.
The evaluation focused on classification performance rather than defect-depth estimation or identification of individual wave components.

The main contributions of this study are as follows.

\begin{itemize}
    \item We constructed FA-PI using an EMPH parameterization in which the birth coordinate corresponds to frequency-bin location and the persistence coordinate is proportional to Fourier amplitude, and derived the analytical relationship between its continuous persistence surface and Gaussian smoothing of the Fourier amplitude spectrum.
        
    \item We demonstrated that FA-PI achieved consistently higher mean performance across the reported classification metrics than both Fourier baselines under the evaluated condition-matched noisy settings for binary classification on the experimental backfill-void dataset.

    \item We further demonstrated the classification benefit of FA-PI relative to both Fourier baselines through separate training and evaluation on a public bridge-deck dataset.
\end{itemize}

\section{Methodology}
\label{sec:methodology}

\subsection{Physical Basis of Impact-Echo Spectral Analysis}

Impact-echo signals are generated by applying a mechanical impact to the surface of a concrete structure, producing transient elastic stress waves that propagate through the material.
When subsurface defects such as voids are present, part of the incident wave is reflected because of the acoustic impedance contrast between the concrete and the medium within the defect.
In particular, the large impedance contrast between concrete and air causes most of the incident wave energy to be reflected at an air-filled interface \cite{carino2015impact}.

Repeated wave reflections between the concrete surface and an internal interface can produce a characteristic peak in the Fourier amplitude spectrum.
For a sufficiently extensive planar reflector approximately parallel to the surface, the corresponding peak frequency can be related to the reflector depth when the impact excites the relevant frequency range \cite{carino2001impact,carino2015impact}.
Under these conditions and for a given wave speed, the peak frequency decreases as the reflector depth increases.

When the structure is sufficiently large and geometrically simple, reflections from the external boundaries have relatively little influence. As a result, the peak associated with waves repeatedly traveling between the surface and an internal defect may appear clearly in the amplitude spectrum.
Under these conditions, the dominant peak frequency can be related to the defect depth relatively easily.

In contrast, when the structure is small or has a complex geometry, waves may be reflected not only by the defect but also by the side boundaries and corners of the structure. In addition, the impact may excite multiple structural vibration modes. Consequently, several spectral peaks arising from different physical mechanisms may appear simultaneously in the amplitude spectrum. Because real defects also have finite lateral dimensions, waves may diffract or scatter at the defect edges, further complicating the interpretation of the spectral response.

\subsection{Construction of the Frequency--Amplitude Persistence Image}
\noindent\textbf{(i) Fourier amplitude spectrum.}
Let $f=(f(0),\ldots,f(T-1))\in\mathbb{R}^{T}$ be a discrete signal.
Its discrete Fourier transform is
\begin{equation}
    \hat{f}(L)
    =
    \frac{1}{T}\sum_{t=0}^{T-1}f(t)e^{-2\pi iLt/T},
    \qquad L=0,\ldots,T-1.
    \label{eq:dft}
\end{equation}
In practice, these coefficients are computed using the fast Fourier transform (FFT) algorithm.
For the first $N$ positive-frequency bins, we define
\begin{equation}
    r_L^f=2|\hat f(L)|,\qquad L=1,\ldots,N,
    \label{eq:amplitude_coefficient}
\end{equation}
and collect them into the amplitude spectrum vector $\mathbf r^f=(r_1^f,\ldots,r_N^f)$.
Here, $L$ denotes the frequency-bin index, and the corresponding physical frequency is $\nu_L=LF_s/T$, where $F_s$ is the sampling frequency.

\noindent\textbf{(ii) Gaussian-smoothed Fourier amplitude spectrum.}
For comparison, we consider Gaussian smoothing of the amplitude spectrum along the frequency axis, following the standard Gaussian convolution
framework \cite{young1995recursive}. Let
\begin{equation}
    G_\sigma(u)
    =
    \frac{1}{\sqrt{2\pi}\sigma}
    \exp\!\left(
        -\frac{u^2}{2\sigma^2}
    \right),
    \label{eq:gaussian_kernel}
\end{equation}
where $\sigma>0$ denotes the Gaussian kernel width in frequency-bin units.
The corresponding continuous Gaussian-smoothed representation is defined as
\begin{equation}
    \widetilde r_\sigma^f(x)
    =
    \sum_{L=1}^{N}
    r_L^f G_\sigma(x-L).
    \label{eq:smoothed_spectrum}
\end{equation}
Here, $x\in\mathbb{R}$ is a continuous coordinate measured in frequency-bin units.
Thus, each spectral amplitude $r_L^f$ is distributed around its original frequency-bin location $L$, with $\sigma$ controlling the extent of smoothing across neighboring frequency bins.
As $\sigma\to0^+$, the Gaussian kernels become increasingly concentrated at the original frequency-bin
locations.

\noindent\textbf{(iii) Frequency--amplitude representation.}
EMPH provides an analytical mapping from the Fourier amplitude spectrum to a persistence diagram.
For the retained spectrum $\mathbf r^f=(r_1^f,\ldots,r_N^f)$, EMPH defines
\begin{equation}
    \mathbf r^f
    \xrightarrow{\;\mathrm{EMPH}_{\ad,\e}\;}
    \dgm_1(f),
    \qquad
    \ad\in\mathbb{R}_{>0}^{N},
    \quad
    \e\in\mathbb{R}^{N}.
    \label{eq:emph_mapping}
\end{equation}
Here, \(\ad\) and \(\e\) are parameter vectors of the EMPH construction that determine how the Fourier components are mapped to the persistence diagram, and \(\dgm_1(f)\) denotes the resulting persistence diagram; see Section 4.3 of \cite{kim2026exact} for the underlying construction.

In this study, we use the following parameterization:
\[
\ad=(1,\ldots,1),\qquad
\e=(-1,-2,\ldots,-N).
\]

We keep this parameterization fixed throughout the experiments.
Under this choice, the EMPH construction yields
\begin{equation}
    \dgm_1(f)
    =
    \left\{
        \left(
            L,
            L+\sqrt{3}\,r_L^f
        \right)
        :
        L=1,\ldots,N
    \right\}.
    \label{eq:canonical_dgm}
\end{equation}
Writing each point as $(b_L,d_L)$ and defining its persistence as $p_L=d_L-b_L$, the corresponding birth--persistence coordinates are
\begin{equation}
    (b_L,p_L)
    =
    \left(
        L,
        \sqrt{3}\,r_L^f
    \right),
    \qquad
    L=1,\ldots,N.
    \label{eq:canonical_coordinates}
\end{equation}

Thus, the birth coordinate equals the retained frequency-bin index, while the persistence coordinate is proportional to the corresponding Fourier amplitude.

\noindent\textbf{(iv) Persistence surface and its relationship to Gaussian smoothing.}
Following the persistence image framework \cite{adams2017persistence}, we construct a continuous persistence surface from the birth--persistence coordinates.
Each point $(b_L,p_L)$ contributes a Gaussian kernel centered at $(b_L,p_L)$ and weighted by $p_L$, giving
\begin{equation}
    \rho_{\sigma}[f](x,y)
    =
    \sum_{L=1}^{N}
    p_L
    \exp\!\left(
        -\frac{
            (x-b_L)^2+(y-p_L)^2
        }{2\sigma^2}
    \right),
    \label{eq:persistence_surface}
\end{equation}
Here, $x$ and $y$ denote continuous birth and persistence coordinates,
respectively.
The Gaussian kernel width $\sigma>0$ controls the spread along both coordinates.

The persistence surface has two direct connections with the amplitude spectrum.
First, at the kernel center $(b_L,p_L)$ associated with the $L$-th retained Fourier component, we have
\begin{equation}
    \lim_{\sigma\to0^+}
    \rho_{\sigma}[f](b_L,p_L)
    =
    p_L
    =
    \sqrt{3}\,r_L^f,
    \qquad
    L=1,\ldots,N.
    \label{eq:zero_bandwidth_recovery}
\end{equation}
Thus, the surface value at each kernel center converges to $\sqrt{3}$ times the corresponding retained Fourier amplitude as $\sigma\to0^+$.

Second, integrating the continuous persistence surface over the persistence coordinate gives a direct relationship with one-dimensional Gaussian smoothing.
Using $b_L=L$ and $p_L=\sqrt{3}r_L^f$, we obtain
\begin{align}
    \int_{-\infty}^{\infty}
    \rho_{\sigma}[f](x,y)\,dy
    &=
    \sqrt{2\pi}\sigma
    \sum_{L=1}^{N}
    p_L
    \exp\!\left(
        -\frac{(x-L)^2}{2\sigma^2}
    \right)
    \notag\\
    &=
    \sqrt{6\pi}\sigma
    \sum_{L=1}^{N}
    r_L^f
    \exp\!\left(
        -\frac{(x-L)^2}{2\sigma^2}
    \right).
    \label{eq:persistence_marginal}
\end{align}
By Eq.~\eqref{eq:smoothed_spectrum},
\begin{equation}
    \boxed{
    \int_{-\infty}^{\infty}
    \rho_{\sigma}[f](x,y)\,dy
    =
    2\sqrt{3}\pi\sigma^2
    \widetilde r_\sigma^f(x)
    }.
    \label{eq:smoothed_spectrum_relation}
\end{equation}
Thus, marginalizing the persistence surface over the persistence coordinate recovers the one-dimensional Gaussian-smoothed amplitude spectrum up to the explicit scale factor $2\sqrt{3}\pi\sigma^2$.

\begin{figure*}[h]
    \centering
    \includegraphics[width=1.0\linewidth]{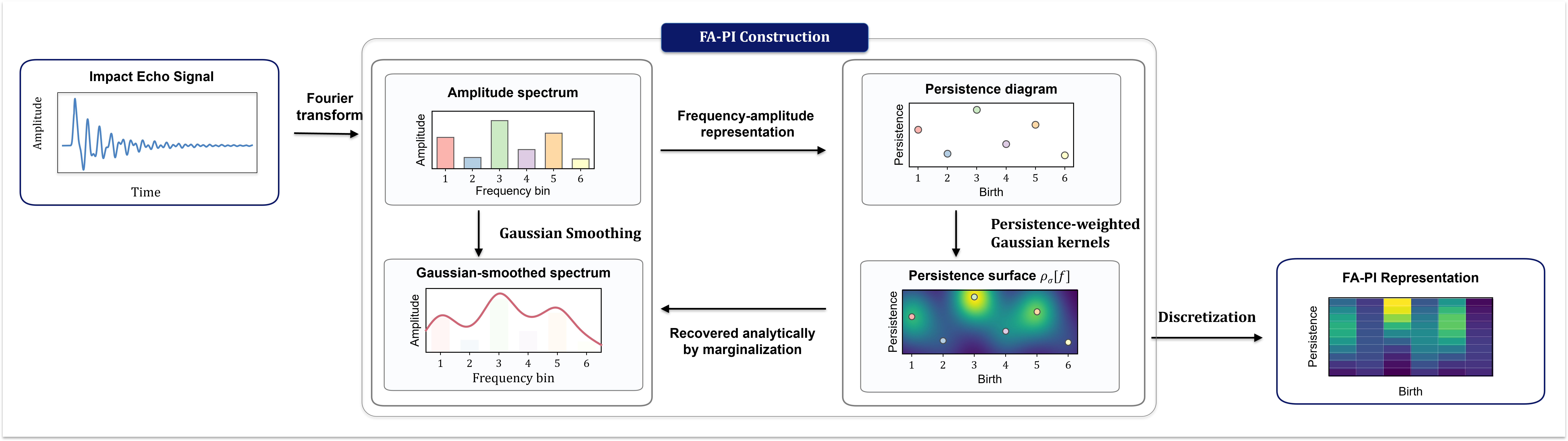}
    \caption{Overview of the frequency--amplitude persistence image (FA-PI) construction from impact-echo signals.
    The retained Fourier components are represented in birth--persistence coordinates, with birth equal to the frequency-bin index and persistence proportional to the corresponding Fourier amplitude.
    Persistence-weighted Gaussian kernels form a continuous persistence surface, which is sampled on a finite grid to obtain FA-PI.
    Marginalizing the continuous surface over the persistence coordinate yields the corresponding one-dimensional Gaussian-smoothed Fourier amplitude spectrum.}
    \label{fig:pipeline}
\end{figure*}

\noindent\textbf{(v) Numerical discretization.}
The Gaussian-smoothed spectrum in Eq.~\eqref{eq:smoothed_spectrum} and the persistence surface in Eq.~\eqref{eq:persistence_surface} are defined as continuous functions to make their analytical relationship explicit. 
For numerical evaluation, these continuous representations are converted to finite-dimensional forms.
For the Gaussian-smoothed Fourier amplitude spectrum, we first apply a normalized discrete Gaussian filter to the full positive-frequency spectrum and then retain the first \(N\) frequency bins. To obtain a finite-dimensional persistence image for classification, the persistence surface is sampled on a uniform rectangular grid over the birth--persistence domain. In our experiments, the rectangular domain
\[
[1,N]\times[0,y_{\max}]
\]
is discretized into an \(R_x\times R_y\) grid with
\[
R_x=N,\qquad R_y=100,
\]
where \(y_{\max}\) is set to the maximum persistence value observed in the training data. Sampling \(\rho_\sigma[f]\) on this grid yields FA-PI, denoted by \(\PI_\sigma[f]\).
The overall FA-PI construction and its relationship to the Fourier amplitude spectrum are summarized in Fig.~\ref{fig:pipeline}.

\section{Datasets and Experimental Setup}
\label{sec:experiments}

We evaluated FA-PI on two impact-echo datasets using original signals and signals with artificially added noise.
Models were trained and evaluated separately within each dataset.
For each noisy condition, the noise type and signal-to-noise ratio (SNR) were matched across training, validation, and testing.

\subsection{Datasets}
\label{subsec:dataset}
Two impact-echo datasets with different specimen configurations and classification settings are considered.

\subsubsection{Experimental Backfill-Void Dataset}

\begin{figure*}[!t]
\centering
\includegraphics[width=\linewidth]{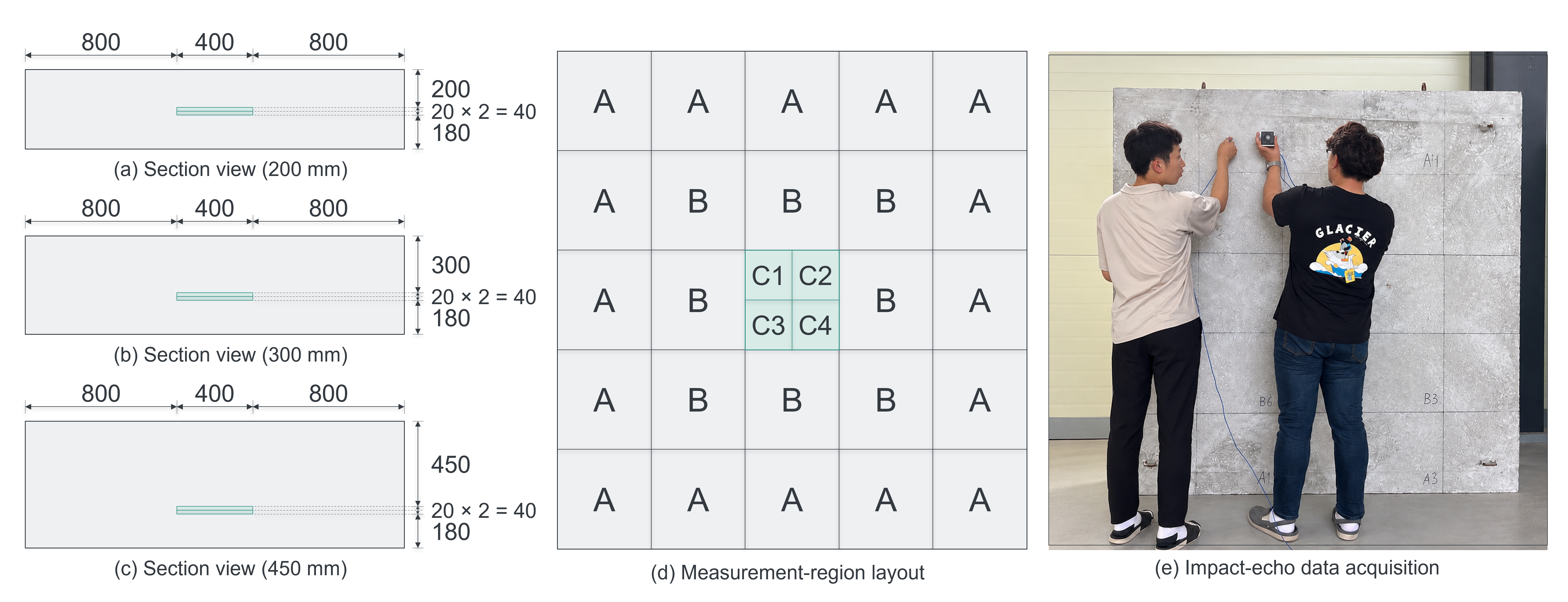}
\caption{Experimental backfill-void specimens and impact-echo data acquisition:
(a)--(c) section views showing embedded Styrofoam inserts at defect depths of 200, 300, and 450~mm, respectively;
(d) measurement-region layout showing regions A, B, and C, with region C subdivided into C1--C4;
and (e) impact-echo data acquisition on a concrete specimen.
Defect depth is measured from the measurement surface to the upper boundary of the Styrofoam insert.
All dimensions are in millimetres.}
\label{fig:specimen}
\end{figure*}

We used the experimental backfill-void dataset previously analyzed in our CWT--CNN study \cite{lee2024void}.
It contains 3,137 impact-echo signals acquired from three concrete specimens with simulated subsurface voids at different depths.
As illustrated in Fig.~\ref{fig:specimen}, Styrofoam was embedded to simulate subsurface voids at defect depths of 200, 300, and 450~mm.
Here, defect depth denotes the distance from the measurement surface to the upper boundary of the simulated void.
Measurements were collected at predefined surface locations belonging to regions A, B, and C, as shown in Fig.~\ref{fig:specimen}(d).
Regions A and B correspond to sound concrete near the specimen boundary and in the interior, respectively.
Region C corresponds to the surface area above the simulated void and is subdivided into C1--C4 in the figure.
In the primary analysis, regions A and B are combined into a single sound-concrete class and classified against region C as a binary task. As a supplementary analysis, we also considered a three-class task in which regions A, B, and C were treated as separate classes.

The numbers of signals for each specimen and measurement region are summarized in Table~\ref{tab:dataset}.
Each signal was sampled at \(F_s=40{,}000\)~Hz and contains \(T=2{,}000\) time samples, corresponding to a frequency-bin spacing of
\[
    \Delta\nu=\frac{F_s}{T}=20~\mathrm{Hz}.
\]

\begin{table}[H]
\centering
\caption{Number of impact-echo signals by defect depth and measurement region.}
\label{tab:dataset}
\begin{tabular}{ccccc}
\hline
Depth (mm) & A & B & C & Total \\
\hline
200 & 200 & 321 & 500 & 1021 \\
300 & 208 & 320 & 500 & 1028 \\
450 & 249 & 329 & 510 & 1088 \\
\hline
Total & 657 & 970 & 1510 & 3137 \\
\hline
\end{tabular}
\end{table}

\subsubsection{Public Bridge-Deck Dataset}
We also used a public bridge-deck impact-echo dataset \cite{dorafshan2020deep,dorafshan2020dataset}.
It contains 2,016 impact-echo signals acquired from eight reinforced concrete bridge-deck specimens constructed in the laboratory.
Of these signals, 1,504 are labeled as sound and 512 as defect.
We used these labels for binary classification.

Impact-echo measurements were recorded at a sampling frequency of
\(F_s=200{,}000\)~Hz over a duration of 10~ms, resulting in
\(T=2{,}000\) time samples per signal. The corresponding frequency-bin
spacing is
\[
    \Delta\nu=\frac{F_s}{T}=100~\mathrm{Hz}.
\]

\subsection{Noise Generation}
\begin{figure*}[!htbp]
\centering
\includegraphics[width=0.85\textwidth]{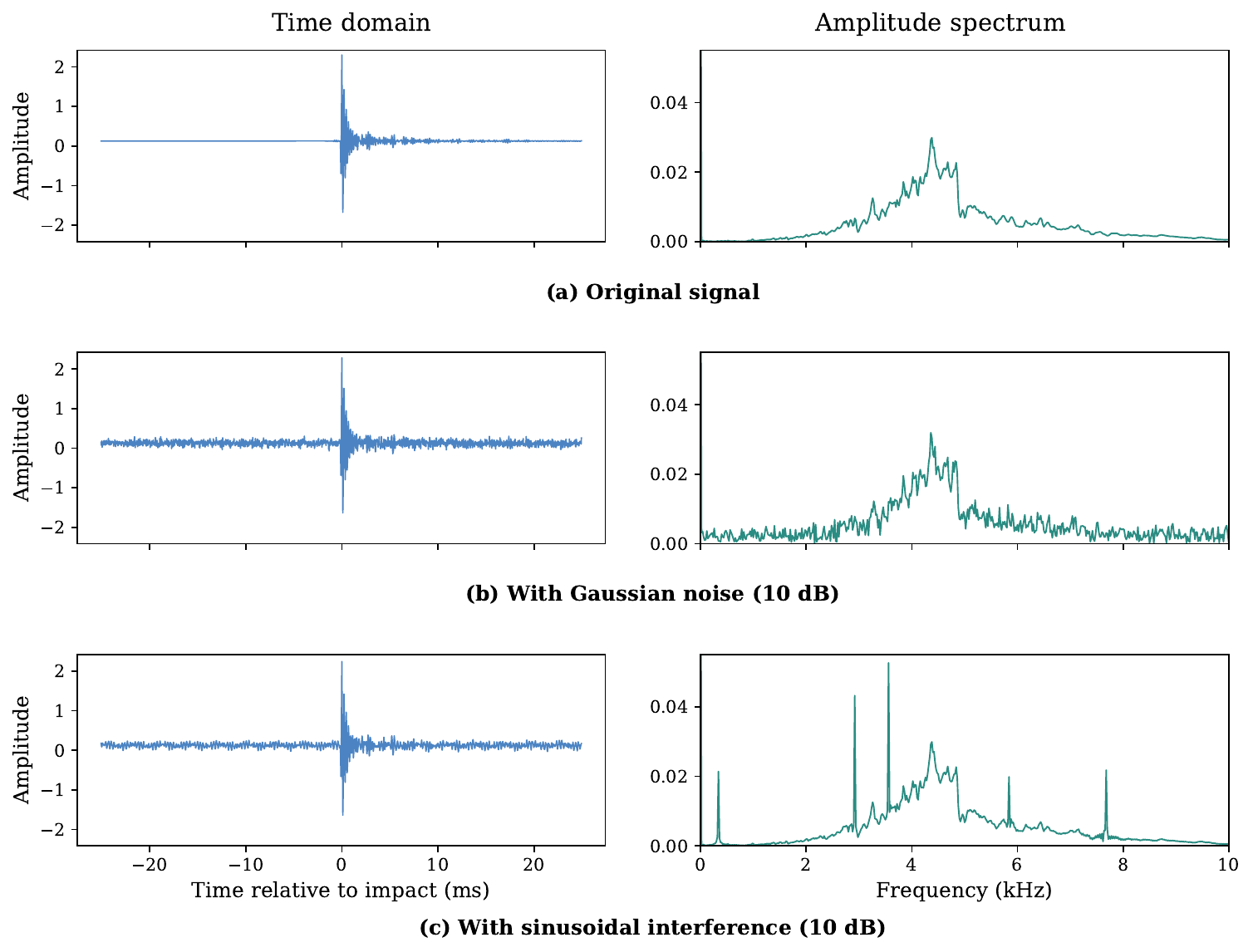}
\caption{Effects of Gaussian noise and sinusoidal interference on a representative impact-echo signal at 10~dB SNR. The left column shows the time-domain signals, and the right column shows the corresponding amplitude spectra. Gaussian noise introduces broadband spectral fluctuations, whereas sinusoidal interference produces localized structured spectral components.}
\label{fig:noise_examples}
\end{figure*}

For both datasets, we evaluated classification performance using signals with artificially added Gaussian noise, sinusoidal interference, or mixed noise.
For each original signal \(f\) and an unscaled noise realization \(q\), the added noise was scaled to a target SNR \(\rho\), expressed in decibels, as
\begin{equation}
    \widetilde q
    =
    \frac{\|f\|_2}{\|q\|_2}
    10^{-\rho/20}q,
\end{equation}
and the resulting noisy signal is
\[
    f^{(\rho)}=f+\widetilde q.
\]
Here, \(\|\cdot\|_2\) denotes the Euclidean norm, and the target SNR is defined relative to the original recorded signal and the added noise.
The target SNR levels were 20, 10, 5, and 0~dB.
The unscaled noise realization \(q\) is defined below for each noise type.
Fig.~\ref{fig:noise_examples} illustrates the effects of Gaussian noise and sinusoidal interference on a representative impact-echo signal at 10~dB SNR.

\paragraph{Gaussian Noise}

Gaussian noise was used as a simplified model of broadband
measurement noise, consistent with previous impact-echo classification studies
that evaluated signals corrupted by additive Gaussian noise
\cite{xiang2002detection}.
The unscaled noise samples were drawn independently from a zero-mean,
unit-variance Gaussian distribution:
\begin{equation}
    q_t^{\mathrm{G}}\sim\mathcal{N}(0,1),
    \qquad t=0,\ldots,T-1.
\end{equation}

\paragraph{Sinusoidal Interference}

In impact-echo measurements, reflections from structural boundaries and the excitation of multiple vibration modes can introduce additional characteristic peaks into the amplitude spectrum \cite{carino2015impact}.
To evaluate classification under synthetic structured spectral interference, we generated the following unscaled sum of sinusoidal components:
\begin{equation}
\begin{aligned}
q_t^{\mathrm{S}}
={}&
\sum_{k=1}^{K}
A_k
\sin\left(
    2\pi m_k\frac{t}{T}+\phi_k
\right) \\
&+
\sum_{k=1}^{K}\sum_{\ell=1}^{J}
A_k c_{k,\ell}
\sin\left(
    2\pi(m_k+\delta_{k,\ell})\frac{t}{T}
    +\phi_{k,\ell}
\right),
\end{aligned}
\label{eq:sinusoidal_interference}
\end{equation}
where \(K\) is the number of main sinusoidal components and \(J\) is the number of additional components near each main frequency.
The parameter \(m_k\) denotes the frequency-bin index of the \(k\)-th main component, and \(A_k\) is its unscaled amplitude.
For each additional component, \(c_{k,\ell}\) is its amplitude relative to the corresponding main component, and \(\delta_{k,\ell}\) is its frequency offset in frequency-bin units.
The phases \(\phi_k\) and \(\phi_{k,\ell}\) are expressed in radians.
The parameter settings are summarized in Table~\ref{tab:noise_parameters}.
This construction introduces controlled spectral interference rather than physically simulating boundary reflections.

\begin{table}[!htbp]
\centering
\caption{Parameter settings for generating sinusoidal interference.}
\label{tab:noise_parameters}
\renewcommand{\arraystretch}{1.1}
\begin{tabular}{cc}
\hline
Parameter & Experimental setting \\
\hline
\(K\) & \(5\) \\
\(J\) & \(4\) \\
\(m_k\) & \(\mathrm{Unif}\{1,\ldots,400\}\), without replacement \\
\(A_k\) & \(\mathcal{U}(1,1.5)\) \\
\(c_{k,\ell}\) & \(\mathcal{U}(0.3,0.6)\) \\
\(\delta_{k,\ell}\) &
\(\mathcal{U}([-0.30,-0.05]\cup[0.05,0.30])\) \\
\(\phi_k,\phi_{k,\ell}\) & \(\mathcal{U}(0,2\pi)\) \\
\hline
\end{tabular}
\end{table}

\paragraph{Mixed Noise}

For the mixed-noise condition, the Gaussian and sinusoidal components were each normalized to unit root-mean-square (RMS) amplitude before being combined.
Denoting the normalized components by \(\bar q^{\mathrm G}\) and \(\bar q^{\mathrm S}\), respectively, the mixed-noise realization was defined as
\begin{equation}
    q^{\mathrm M}=\bar q^{\mathrm S}+\bar q^{\mathrm G}.
\end{equation}
The combined realization was then scaled to the target SNR using the procedure described above.

\subsection{Experimental Setup}

\subsubsection{Evaluated Representations and Classifiers}

We compared the raw Fourier amplitude spectrum, the Gaussian-smoothed Fourier amplitude spectrum, and FA-PI using the same type of linear classifier with a Softmax output.
This comparison assessed the classification performance of the two-dimensional frequency--amplitude representation relative to the raw spectrum and frequency-axis smoothing.

CWT--Softmax was included as an additional time--frequency benchmark.
Each signal was transformed using the generalized Morse wavelet \citep{lilly2012generalized}, and the magnitude of the CWT coefficients was normalized by its maximum value for that signal.
The normalized CWT magnitude was resized to \(224\times224\) using bicubic interpolation and vectorized for input to a linear classifier with a Softmax output.

The raw Fourier amplitude spectrum, the Gaussian-smoothed Fourier amplitude spectrum, FA-PI, and the CWT magnitude representation were evaluated using the same type of linear classifier with a Softmax output.
In the primary experiments, the number of retained positive-frequency bins was set to \(N=500\) for all three Fourier-derived representations.
Retaining \(N=500\) bins corresponded to upper frequencies of 10~kHz for the experimental backfill-void dataset and 50~kHz for the public bridge-deck dataset because the analyzed signal durations differed.
Thus, the same number of retained bins did not imply the same physical frequency cutoff across the two datasets.

\subsubsection{Data Splitting, Model Selection, and Evaluation Settings}
Each impact-echo signal was treated as a single sample.
For each dataset, these samples were assigned to training and held-out test sets using a stratified 70:30 split.
Five-fold stratified cross-validation was performed within the training set.
Within each classification task, all methods used identical training/test partitions and cross-validation folds.
The test set was excluded from model selection.

For each noisy condition, the noise type and SNR level were matched across training, validation, and testing.
Gaussian noise, sinusoidal interference, and mixed noise were evaluated at four SNR levels.
This yielded 12 noisy conditions per classification task and 24 across the two primary binary classification tasks, excluding the original-signal conditions.

Models were trained using cross-entropy loss.
For Smoothed-FFT--Softmax, the Gaussian kernel width was selected from \(\sigma\in\{0.25,0.5,1,2,4\}\) frequency bins.
For FA-PI--Softmax, the Gaussian kernel width was selected from \(\sigma\in\{0.005,0.01,0.05,0.1,0.5,1\}\).
For each candidate width, the checkpoint with the lowest validation loss was retained within each fold.
The corresponding validation losses were averaged across the five folds, and the width with the lowest mean validation loss was selected.
The selected epoch could differ between folds.

To assess sensitivity to spectral truncation, we repeated the comparison of FA-PI with the raw and Gaussian-smoothed Fourier amplitude spectra for \(N\in\{200,300,400,500,600,1000\}\).
For each value of \(N\), Gaussian kernel widths were selected using the same validation-based procedure.

To assess sensitivity to the FA-PI Gaussian kernel width, we fixed \(N=500\) and evaluated FA-PI--Softmax at each candidate width \(\sigma\in\{0.005,0.01,0.05,\allowbreak 0.1,0.5,1\}\) on the two binary classification tasks.
This analysis evaluated all candidate widths rather than only the validation-selected width.
For each width, test accuracy was first averaged across the five fold-specific models within each condition and then across the original-signal condition and the 12 condition-matched noisy conditions.
This analysis was used only to assess sensitivity; the primary results used widths selected from validation data.

Classification performance was evaluated using accuracy and macro-averaged precision, recall, and F1-score.
For each evaluation condition, these metrics were calculated separately for the five fold-specific models on the same held-out test set and reported as the mean and standard deviation across models.
The main text presents summaries of all four metrics for the two binary classification tasks.
The supplementary analysis additionally reports the corresponding results for the three-class task on the experimental backfill-void dataset.
Detailed condition-specific results, including fold standard deviations, are provided in Tables~\ref{tab:supp_fig5}--\ref{tab:supp_f1} of the Supplementary Material.

\section{Results}
\label{sec:results}

\subsection{Classification Performance under Original and Noisy Conditions}

\begin{figure*}[!htbp]
\centering
\includegraphics[width=\textwidth]{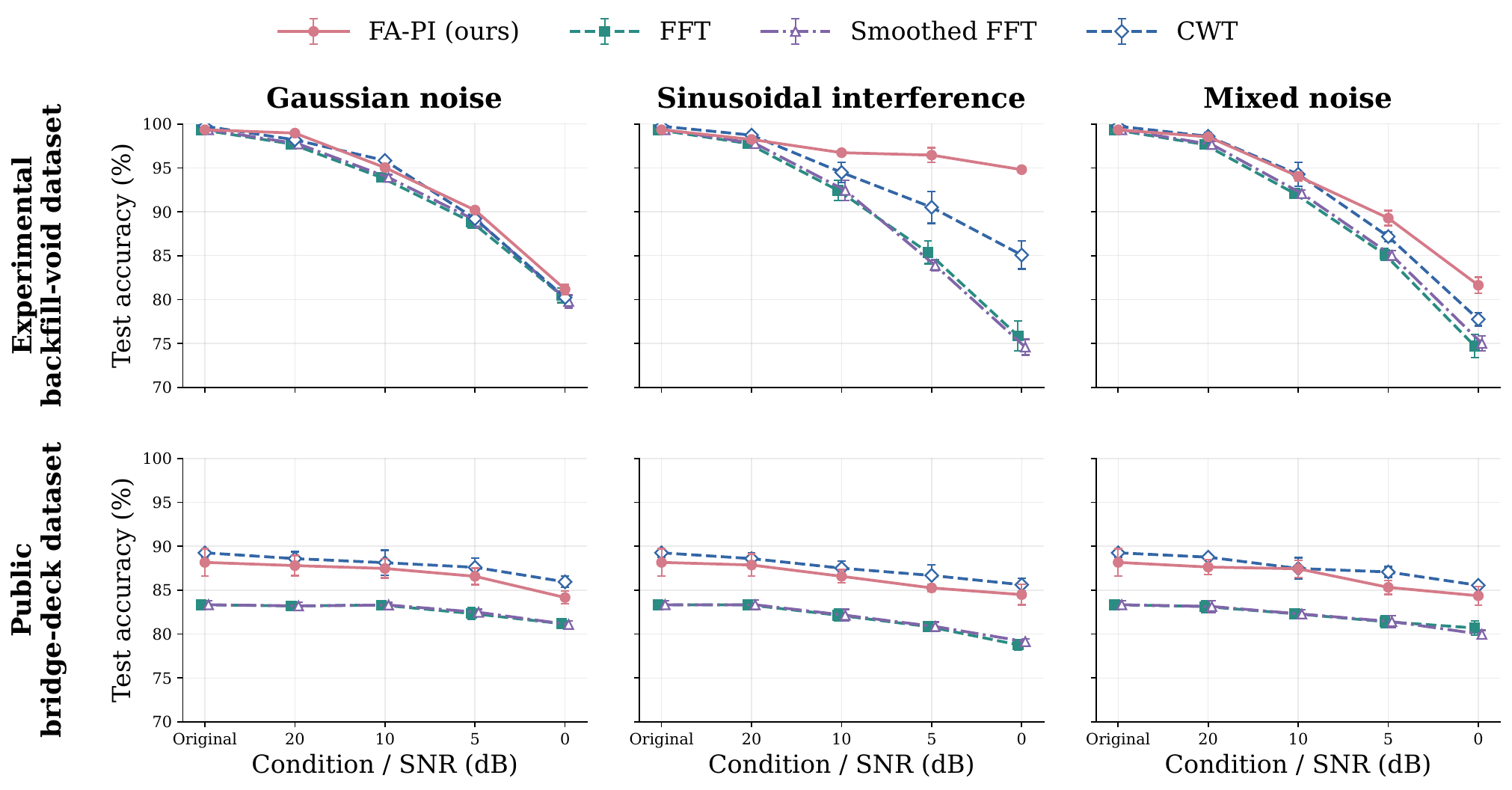}
\caption{Binary classification performance on original signals and under condition-matched noisy settings.
Rows correspond to the experimental backfill-void dataset and the public bridge-deck dataset, respectively.
Columns correspond to Gaussian noise, sinusoidal interference, and mixed noise.
Each point shows mean test accuracy across five fold-specific models, and error bars indicate the corresponding standard deviation.
For each noisy condition, the noise type and SNR were matched across training, validation, and testing.}
\label{fig:matched_results}
\end{figure*}

Tables~\ref{tab:backfill_binary_summary_draft} and~\ref{tab:bridge_deck_binary_summary} summarize accuracy and macro-averaged precision, recall, and F1-score on the experimental backfill-void dataset and the public bridge-deck dataset, respectively.
Original-signal results are means across five fold-specific models, while noise-type summaries additionally average over the four
evaluated SNR levels.
Figure~\ref{fig:matched_results} presents the condition-specific mean test accuracies, with error bars indicating the standard
deviation across the five fold-specific models within each condition.
Detailed condition-specific results are provided in Tables~\ref{tab:supp_fig5}--\ref{tab:supp_f1} of the Supplementary Material.

\begin{table}[H]
\centering
\caption{Binary classification performance (\%) on the experimental backfill-void dataset.
Values are averaged across five fold-specific models and, for each noise type, equally across SNR levels of 20, 10, 5, and 0~dB.
Bold indicates the highest mean for each metric within each block, not statistical significance.}
\label{tab:backfill_binary_summary_draft}
\small
\setlength{\tabcolsep}{4.5pt}
\renewcommand{\arraystretch}{1.16}

\begin{tabular*}{\linewidth}{@{\extracolsep{\fill}}lrrrr@{}}
\toprule
Method
& Accuracy
& \shortstack{Macro-\\precision}
& \shortstack{Macro-\\recall}
& \shortstack{Macro-\\F1-score} \\
\midrule

\multicolumn{5}{@{}l}{\textbf{Original signals}} \\
\addlinespace[2pt]
FA-PI--Softmax
& 99.3 & 99.3 & 99.3 & 99.3 \\
FFT--Softmax
& 99.3 & 99.4 & 99.3 & 99.3 \\
Smoothed-FFT--Softmax
& 99.3 & 99.4 & 99.3 & 99.3 \\
CWT--Softmax
& \textbf{99.7} & \textbf{99.7} & \textbf{99.7} & \textbf{99.7} \\
\midrule

\multicolumn{5}{@{}l}{\textbf{Gaussian noise}} \\
\addlinespace[2pt]
FA-PI--Softmax
& \textbf{91.3} & \textbf{91.4} & \textbf{91.3} & \textbf{91.3} \\
FFT--Softmax
& 90.2 & 90.3 & 90.2 & 90.2 \\
Smoothed-FFT--Softmax
& 90.1 & 90.2 & 90.0 & 90.0 \\
CWT--Softmax
& 90.9 & 90.9 & 90.8 & 90.8 \\
\midrule

\multicolumn{5}{@{}l}{\textbf{Sinusoidal interference}} \\
\addlinespace[2pt]
FA-PI--Softmax
& \textbf{96.6} & \textbf{96.6} & \textbf{96.5} & \textbf{96.5} \\
FFT--Softmax
& 87.9 & 87.9 & 87.8 & 87.8 \\
Smoothed-FFT--Softmax
& 87.2 & 87.3 & 87.1 & 87.1 \\
CWT--Softmax
& 92.2 & 92.3 & 92.2 & 92.2 \\
\midrule

\multicolumn{5}{@{}l}{\textbf{Mixed noise}} \\
\addlinespace[2pt]
FA-PI--Softmax
& \textbf{90.9} & \textbf{91.0} & \textbf{90.8} & \textbf{90.8} \\
FFT--Softmax
& 87.4 & 87.5 & 87.3 & 87.3 \\
Smoothed-FFT--Softmax
& 87.4 & 87.5 & 87.4 & 87.4 \\
CWT--Softmax
& 89.4 & 89.6 & 89.4 & 89.4 \\
\bottomrule
\end{tabular*}
\end{table}

\begin{table}[H]
\centering
\caption{Binary classification performance (\%) on the public bridge-deck dataset.
Values are averaged across five fold-specific models and, for each noise type, equally across SNR levels of 20, 10, 5, and 0~dB.
Bold indicates the two best-performing methods for each metric within each block, not statistical significance.}
\label{tab:bridge_deck_binary_summary}
\small
\setlength{\tabcolsep}{4pt}
\renewcommand{\arraystretch}{1.16}

\begin{tabular*}{\linewidth}{@{\extracolsep{\fill}}lrrrr@{}}
\toprule
Method
& Accuracy
& \shortstack{Macro-\\precision}
& \shortstack{Macro-\\recall}
& \shortstack{Macro-\\F1-score} \\
\midrule

\multicolumn{5}{@{}l}{\textbf{Original signals}} \\
\addlinespace[2pt]
FA-PI--Softmax
& \textbf{88.2} & \textbf{88.3} & \textbf{79.5} & \textbf{82.5} \\
FFT--Softmax
& 83.3 & 85.1 & 69.2 & 72.5 \\
Smoothed-FFT--Softmax
& 83.3 & 85.1 & 69.2 & 72.5 \\
CWT--Softmax
& \textbf{89.3} & \textbf{90.5} & \textbf{80.6} & \textbf{84.0} \\
\midrule

\multicolumn{5}{@{}l}{\textbf{Gaussian noise}} \\
\addlinespace[2pt]
FA-PI--Softmax
& \textbf{86.5} & \textbf{86.4} & \textbf{76.5} & \textbf{79.5} \\
FFT--Softmax
& 82.5 & 84.6 & 67.4 & 70.4 \\
Smoothed-FFT--Softmax
& 82.5 & 84.8 & 67.4 & 70.4 \\
CWT--Softmax
& \textbf{87.6} & \textbf{90.1} & \textbf{76.9} & \textbf{80.6} \\
\midrule

\multicolumn{5}{@{}l}{\textbf{Sinusoidal interference}} \\
\addlinespace[2pt]
FA-PI--Softmax
& \textbf{86.0} & \textbf{85.6} & \textbf{75.9} & \textbf{78.9} \\
FFT--Softmax
& 81.3 & 81.3 & 65.7 & 68.2 \\
Smoothed-FFT--Softmax
& 81.4 & 81.9 & 65.7 & 68.2 \\
CWT--Softmax
& \textbf{87.1} & \textbf{88.8} & \textbf{76.4} & \textbf{80.0} \\
\midrule

\multicolumn{5}{@{}l}{\textbf{Mixed noise}} \\
\addlinespace[2pt]
FA-PI--Softmax
& \textbf{86.2} & \textbf{86.2} & \textbf{75.8} & \textbf{78.9} \\
FFT--Softmax
& 81.9 & 83.2 & 66.5 & 69.2 \\
Smoothed-FFT--Softmax
& 81.7 & 82.9 & 66.2 & 68.8 \\
CWT--Softmax
& \textbf{87.2} & \textbf{89.8} & \textbf{76.2} & \textbf{80.0} \\
\bottomrule
\end{tabular*}
\end{table}

On the experimental backfill-void dataset
(Table~\ref{tab:backfill_binary_summary_draft}), all four methods achieved
near-ceiling performance on the original signals.
CWT--Softmax achieved the highest mean accuracy at 99.7\%, while
FA-PI--Softmax, FFT--Softmax, and Smoothed-FFT--Softmax each achieved 99.3\%.

Under the noisy conditions, clearer differences emerged.
FA-PI--Softmax achieved the highest mean accuracy, macro-averaged
precision, recall, and F1-score in each of the three noise-type summaries.
Its mean accuracies were 91.3\%, 96.6\%, and 90.9\% under Gaussian noise,
sinusoidal interference, and mixed noise, respectively, compared with
90.2\%, 87.9\%, and 87.4\% for FFT--Softmax and
90.9\%, 92.2\%, and 89.4\% for CWT--Softmax.
The advantage over the Fourier baselines was most pronounced under sinusoidal interference.

Across the 12 individual noisy conditions shown in
Fig.~\ref{fig:matched_results}, FA-PI--Softmax achieved higher mean
accuracy than both Fourier baselines in all 12 conditions and higher
mean accuracy than CWT--Softmax in 8 conditions.

On the public bridge-deck dataset (Table~\ref{tab:bridge_deck_binary_summary}), FA-PI--Softmax also achieved higher mean accuracy and macro-averaged F1-score than both Fourier baselines for the original signals and
in all three noise-type summaries.
For the original signals, its accuracy and macro-averaged F1-score were 88.2\% and 82.5\%, respectively, compared with
83.3\% and 72.5\% for both Fourier baselines.

CWT--Softmax nevertheless achieved the highest values of all four reported
metrics in every block of the public-dataset summary table.
The relative performance of FA-PI--Softmax and CWT--Softmax therefore
depended on both the dataset and the evaluation condition.
Across the two datasets, the raw and Gaussian-smoothed Fourier baselines
produced similar results.
FA-PI--Softmax consistently outperformed both Fourier baselines in all three noise-type summaries on both datasets and also on the original signals of the public bridge-deck dataset, whereas the original-signal experimental task showed near-ceiling performance with little difference among the Fourier-derived representations.

\subsection{Sensitivity to the Number of Retained Positive-Frequency Bins}
Table~\ref{tab:n_sensitivity} reports mean test accuracy averaged over the
original-signal condition and the 12 condition-matched noisy conditions
for different values of $N$.
FA-PI--Softmax achieved higher mean accuracy than both Fourier baselines
at every evaluated value of $N$ on both datasets.

\begin{table}[!htbp]
\centering
\caption{Mean test accuracy (\%) for different numbers of retained positive-frequency bins \(N\).
Values are averaged across five fold-specific models within each condition and then equally across the original-signal condition and 12 condition-matched noisy conditions.
Gaussian kernel widths were selected using validation loss for each \(N\) and condition.}
\label{tab:n_sensitivity}
\small
\setlength{\tabcolsep}{5pt}
\renewcommand{\arraystretch}{1.16}

\begin{tabular*}{\linewidth}{@{\extracolsep{\fill}}crrrrrr@{}}
\toprule
& \multicolumn{3}{c}{\shortstack{Experimental backfill-void\\dataset}}
& \multicolumn{3}{c}{\shortstack{Public bridge-deck\\dataset}} \\
\cmidrule(lr){2-4}\cmidrule(l){5-7}
\(N\) & FA-PI & FFT & Smoothed FFT
      & FA-PI & FFT & Smoothed FFT \\
\midrule
200  & 91.63 & 86.41 & 86.31
     & 86.78 & 81.89 & 81.83 \\
300  & 93.17 & 88.71 & 88.77
     & 86.82 & 82.01 & 81.98 \\
400  & 93.47 & 89.06 & 88.91
     & 86.47 & 81.92 & 81.94 \\
500  & 93.41 & 89.34 & 89.08
     & 86.40 & 82.01 & 81.98 \\
600  & 93.46 & 89.31 & 89.13
     & 86.49 & 82.05 & 82.01 \\
1000 & 93.31 & 89.18 & 89.00
     & 86.42 & 81.86 & 81.84 \\
\bottomrule
\end{tabular*}
\end{table}

On the experimental backfill-void dataset, FA-PI--Softmax achieved
91.63\% accuracy at $N=200$ and between 93.17\% and 93.47\% for
$N=300$--1000.
In particular, the variation across $N=300$--1000 was only
0.30 percentage points.
On the public bridge-deck dataset, FA-PI--Softmax ranged from
86.40\% to 86.82\% across all evaluated values of $N$.
Thus, the advantage of FA-PI over the Fourier baselines was maintained
throughout the evaluated range and was not specific to the primary
choice of $N=500$.
Increasing $N$ beyond 300 did not produce a consistent improvement on
either dataset.

\subsection{Sensitivity to the Gaussian Kernel Width}
\label{sec:bandwidth_sensitivity}

\begin{figure}[!htbp]
    \centering
    \includegraphics[width=\linewidth]
    {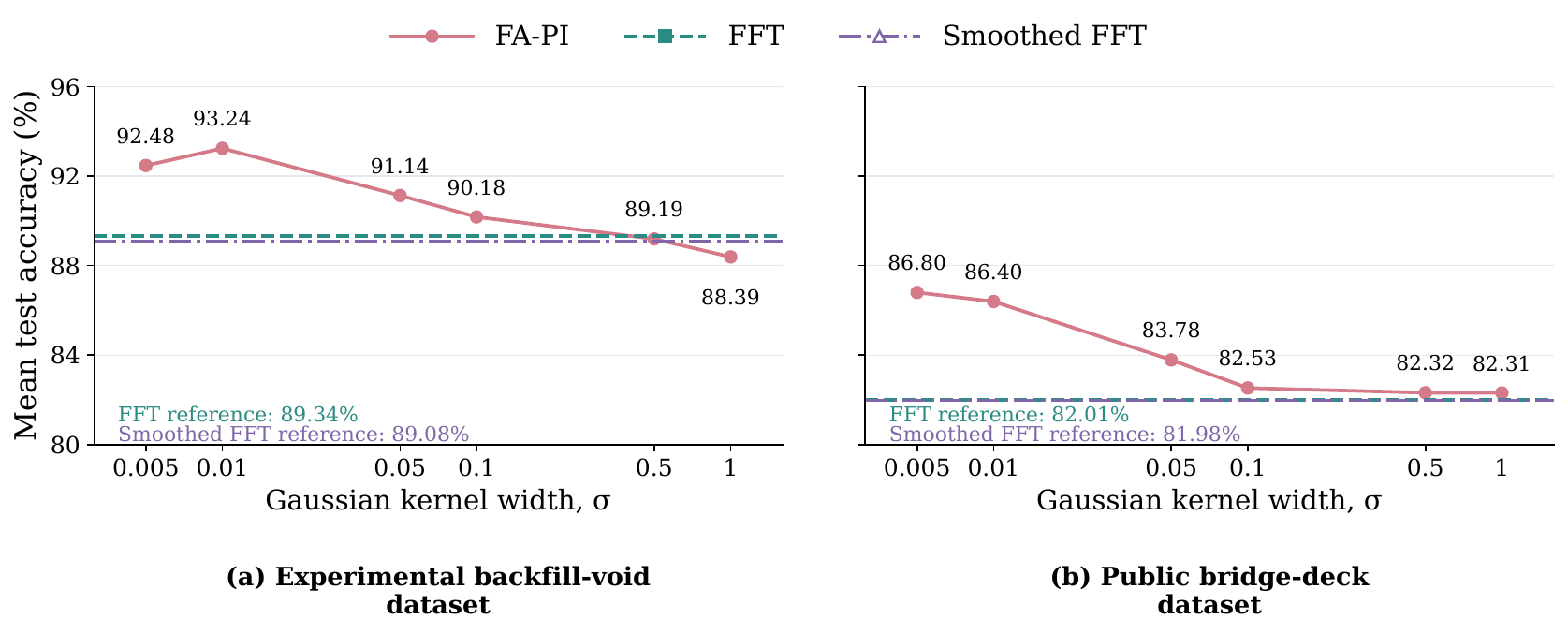}
    \caption{
    Sensitivity of FA-PI--Softmax to the Gaussian kernel width \(\sigma\) at \(N=500\).
    Panels (a) and (b) correspond to the experimental backfill-void dataset and the public bridge-deck dataset, respectively.
    Each point shows mean test accuracy, first averaged across five fold-specific models within each condition and then equally across the original-signal condition and the 12 condition-matched noisy conditions.
    Horizontal lines indicate the corresponding FFT--Softmax and Smoothed-FFT--Softmax reference accuracies at \(N=500\).
    These reference values do not vary with the FA-PI kernel width; Smoothed-FFT--Softmax uses its own validation-selected smoothing width.
    The horizontal axis is logarithmic.
    }
    \label{fig:bandwidth_sensitivity}
\end{figure}

Figure~\ref{fig:bandwidth_sensitivity} shows the sensitivity of
FA-PI--Softmax to the Gaussian kernel width $\sigma$ at $N=500$.
As in the $N$-sensitivity analysis, the reported accuracies were averaged
over the original-signal condition and the 12 condition-matched noisy
conditions.

On the experimental backfill-void dataset, mean accuracy was highest at
$\sigma=0.01$ (93.24\%) and remained above both Fourier baselines for
all evaluated widths up to $\sigma=0.1$.
The accuracy decreased to 88.39\% at $\sigma=1$, giving a range of
4.85 percentage points across the evaluated widths.
On the public bridge-deck dataset, mean accuracy ranged from 86.80\% at
$\sigma=0.005$ to 82.31\% at $\sigma=1$, but remained above both
Fourier baselines at every evaluated width.
Thus, although performance varied with the kernel width, the advantage
of FA-PI over the Fourier baselines was retained over a substantial
portion of the evaluated range of kernel widths and was not confined to a
single narrowly tuned value.
Smaller kernel widths generally provided better performance within the
evaluated range.

These fixed-width results should be distinguished from the primary
comparison, in which the kernel width was selected separately for each
condition using validation loss.
Under that validation-based procedure, the mean accuracies averaged
across the original-signal condition and the 12 noisy conditions were
93.4\% and 86.4\% for the experimental backfill-void and public
bridge-deck datasets, respectively.
The fixed-width test results were used only to assess sensitivity and
were not used to select the kernel width for the primary comparison.

\section{Discussion}
\label{sec:Discussion}

The results indicate that FA-PI can improve impact-echo defect classification relative to raw and Gaussian-smoothed Fourier amplitude spectra within the evaluated settings.
Its performance relative to CWT--Softmax, however, depended on the dataset.

\subsection{Beyond One-Dimensional Frequency Smoothing}

In the evaluated settings, Gaussian smoothing along the frequency axis produced classification performance close to that of the raw Fourier amplitude spectrum.
FA-PI instead represents frequency-bin location through the birth coordinate and Fourier amplitude through a proportional persistence coordinate.
Thus, amplitude affects not only where each Gaussian contribution is placed in the two-dimensional representation but also how strongly it contributes.
This construction changes the representation supplied to the classifier without introducing additional measured information.

The analytical relationship in Eq.~\eqref{eq:smoothed_spectrum_relation} makes the connection to frequency-axis smoothing explicit.
Marginalizing the continuous persistence surface over the entire persistence coordinate yields the corresponding one-dimensional Gaussian-smoothed Fourier amplitude spectrum up to a known scale factor.
This identity applies to the continuous surface and does not imply exact recovery by summing pixels of the finite FA-PI.
It establishes a relationship between the two constructions, but does not by itself explain the observed classification improvement.

\subsection{Dataset-Dependent Performance and Parameter Sensitivity}

On the experimental backfill-void dataset, FA-PI--Softmax achieved higher
accuracy and macro-averaged F1-score than CWT--Softmax in the three noisy
condition summaries, whereas CWT--Softmax performed slightly better on the
original signals.
On the public bridge-deck dataset, however, CWT--Softmax achieved higher values of all four reported metrics.
Unlike FA-PI, which is constructed from the global Fourier amplitude spectrum, CWT retains temporal localization.
This difference may partly explain the dataset-dependent performance, although its contribution was not isolated in the present experiments.
The improvement over the Fourier baselines therefore does not imply a universal advantage over time--frequency representations.

The sensitivity analyses further define the scope of the observed improvement.
When accuracy was averaged across the original-signal condition and the 12 condition-matched noisy conditions, FA-PI remained above both Fourier baselines at every evaluated value of \(N\).
The advantage was therefore not restricted to the primary setting of \(N=500\).
Sensitivity to the Gaussian kernel width was observed on both datasets, with performance generally decreasing at larger widths.
These findings support selecting the kernel width by validation rather than assuming that a single width is suitable across datasets.

\subsection{Limitations and Future Directions}

The evaluation used signal-level rather than specimen-level data splits.
It therefore assessed classification within each dataset, not generalization to previously unseen specimens.
Models were also trained and evaluated separately for the two datasets, so the public-dataset evaluation does not demonstrate transfer of a model trained on the experimental dataset.

The noisy evaluations matched the noise type and SNR across training, validation, and testing.
They do not establish performance under unseen noise conditions or training--test noise mismatch.
The reported standard deviations describe variation among five fold-specific models evaluated on the same held-out test set, rather than variability across independently held-out specimens.

Using the same type of linear classifier did not equalize input dimensionality or the number of trainable parameters across representations.
The present comparison therefore does not isolate the contributions of the coordinate mapping, amplitude-dependent weighting, and Gaussian kernels.
Ablation analyses and comparisons that control input dimensionality would help clarify these contributions.

Further work could incorporate temporal localization into the frequency--amplitude representation.
Another direction is to use separate Gaussian kernel widths along the birth and persistence coordinates, allowing their effects to be examined independently.

\section{Conclusion}
\label{sec:conclusion}
This study constructed FA-PI, a two-dimensional frequency--amplitude representation for impact-echo defect classification.
Under the EMPH parameterization used in this study, the birth coordinate corresponds to the retained frequency-bin index, while the persistence coordinate is proportional to the corresponding Fourier amplitude.
We further established a direct analytical relationship between the continuous persistence surface and the corresponding Gaussian-smoothed Fourier amplitude spectrum.

Using the same type of linear classifier with a Softmax output, FA-PI achieved
higher mean test accuracy than the raw and Gaussian-smoothed Fourier baselines
under most evaluated condition-matched noisy settings across the two datasets.
On the original signals, the three Fourier-derived representations achieved
near-ceiling and nearly identical accuracy on the experimental backfill-void
dataset, whereas FA-PI achieved higher mean test accuracy than both Fourier
baselines on the public bridge-deck dataset.

Its relative performance against CWT--Softmax depended on the dataset and
evaluation condition.
Overall, these findings indicate that organizing Fourier spectral information
in a two-dimensional frequency--amplitude representation can improve
Fourier-based impact-echo defect classification within the evaluated settings,
without establishing a universal advantage over time--frequency representations.

\section*{CRediT authorship contribution statement}
\textbf{Keunsu Kim:} Methodology, Software, Validation,
Formal analysis, Investigation, Writing -- original draft,
Writing -- review \& editing.

\textbf{Suyoung Choi:} Conceptualization, Resources, Supervision,
Project administration, Writing -- review \& editing.

\textbf{Meiyan Kang:} Conceptualization, Validation, Investigation,
Data curation, Visualization, Writing -- original draft,
Writing -- review \& editing.

\section*{Declaration of competing interest}
The authors report no competing financial interests or personal relationships relevant to this work.

\section*{Acknowledgments}
This research was partially supported by JST Moonshot R\&D under Grant Number JPMJMS2021.
S. Choi was supported by the National Research Foundation of Korea Grant funded by the Korean Government (RS-2025-25412264).

\section*{Data availability}

The public bridge-deck impact-echo dataset is available
from Mendeley Data at
\url{https://doi.org/10.17632/44rb96872r.1}
\cite{dorafshan2020dataset}.
The experimental backfill-void dataset was previously used
in our CWT--CNN study \cite{lee2024void} and is not publicly
available because the authors do not have permission to
release it publicly.
Requests for access may be directed to the corresponding authors
and will be considered subject to approval from the data provider.
The implementation code is available at
\url{https://github.com/KeunsuKim/FA-PI-ImpactEcho}.

\section*{Declaration of generative AI and AI-assisted technologies in the manuscript preparation process}

The authors used ChatGPT (OpenAI) and Claude (Anthropic) to assist with literature searches, manuscript organization, English-language revision, and the preparation and revision of data-visualization code. The authors checked and revised the AI-assisted material and accept full responsibility for the final manuscript and the reported results.

\bibliographystyle{elsarticle-num-names}
\bibliography{bibtex/bib/references}

@techreport{sansalone1986impact,
  title       = {Impact-echo: A method for flaw detection in concrete using transient stress waves},
  author      = {Sansalone, Mary and Carino, Nicholas J.},
  institution = {National Bureau of Standards},
  number      = {NBSIR 86-3452},
  address     = {Gaithersburg, MD},
  year        = {1986},
  doi         = {10.6028/NBS.IR.86-3452}
}

@inproceedings{carino2001impact,
  title     = {The impact-echo method: An overview},
  author    = {Carino, Nicholas J.},
  booktitle = {Structures 2001: A Structural Engineering Odyssey},
  editor    = {Chang, Peter C.},
  publisher = {American Society of Civil Engineers},
  address   = {Reston, VA},
  pages     = {1--18},
  year      = {2001},
  doi       = {10.1061/40558(2001)15}
}

@inproceedings{carino2015impact,
  title     = {Impact echo: The fundamentals},
  author    = {Carino, Nicholas J.},
  booktitle = {Proceedings of the International Symposium Non-Destructive Testing in Civil Engineering (NDT-CE)},
  address   = {Berlin, Germany},
  year      = {2015},
  url       = {https://www.ndt.net/article/ndtce2015/papers/257_carino_nicholas.pdf}
}

@book{krautkramer2013ultrasonic,
  title     = {Ultrasonic Testing of Materials},
  author    = {Krautkr{\"a}mer, Josef and Krautkr{\"a}mer, Herbert},
  edition   = {4},
  publisher = {Springer},
  address   = {Berlin, Heidelberg},
  year      = {2013},
  doi       = {10.1007/978-3-662-10680-8}
}

@article{dorafshan2020deep,
  title   = {Deep learning models for bridge deck evaluation using impact echo},
  author  = {Dorafshan, Sattar and Azari, Hoda},
  journal = {Construction and Building Materials},
  volume  = {263},
  pages   = {120109},
  year    = {2020},
  doi     = {10.1016/j.conbuildmat.2020.120109}
}

@article{sanostudy,
  title   = {Study on Applicability of Detection of Internal Cracks by Impact Echo Using Machine Learning},
  author  = {Sano, Shunsuke and Mizobuchi, Toshiaki and Shimbo, Hiroshi and Ozeki, Tomoko and Nojima, Jun-ichiro},
  journal = {e-Journal of Nondestructive Testing},
  volume  = {27},
  number  = {9},
  year    = {2022},
  doi     = {10.58286/27260}
}

@article{wu2025grouting,
  title   = {Grouting defect detection of bridge tendon ducts using impact echo and deep learning via a two-stage strategy},
  author  = {Wu, Yanqi and Huang, Zhengrong and Zhang, Jian and Zhang, Xinyu},
  journal = {Mechanical Systems and Signal Processing},
  volume  = {235},
  pages   = {112955},
  year    = {2025},
  doi     = {10.1016/j.ymssp.2025.112955}
}

@article{lee2025machine,
  title   = {Machine Learning Assisted Method for Automated Impact-Echo Testing of Concrete Structures},
  author  = {Lee, Sang Min and Hong, Jinyoung and Choi, Hajin and Kang, Thomas H.-K.},
  journal = {Journal of Nondestructive Evaluation},
  volume  = {44},
  number  = {4},
  pages   = {121},
  year    = {2025},
  doi     = {10.1007/s10921-025-01260-3}
}

@article{kim2026exact,
  title   = {Exact multi-parameter persistent homology of time-series data: Fast and variable topological inferences},
  author  = {Kim, Keunsu and Jung, Jae-Hun},
  journal = {Advances in Applied Mathematics},
  volume  = {174},
  pages   = {103023},
  year    = {2026},
  doi     = {10.1016/j.aam.2025.103023}
}

@article{adams2017persistence,
  title   = {Persistence images: A stable vector representation of persistent homology},
  author  = {Adams, Henry and Emerson, Tegan and Kirby, Michael and Neville, Rachel and Peterson, Chris and Shipman, Patrick and Chepushtanova, Sofya and Hanson, Eric and Motta, Francis and Ziegelmeier, Lori},
  journal = {Journal of Machine Learning Research},
  volume  = {18},
  number  = {8},
  pages   = {1--35},
  year    = {2017},
  url     = {https://www.jmlr.org/papers/v18/16-337.html}
}

@article{lee2024void,
  title   = {Void detection for tunnel lining backfill using impact-echo method based on continuous wavelet transform and convolutional neural network},
  author  = {Lee, Jiyun and Kim, Kyuwon and Kang, Meiyan and Hong, Eun-Soo and Choi, Suyoung},
  journal = {Geomechanics and Engineering},
  volume  = {36},
  number  = {1},
  pages   = {1--8},
  year    = {2024},
  doi     = {10.12989/gae.2024.36.1.001}
}

@misc{dorafshan2020dataset,
  title        = {Annotated Impact Echo Dataset (bare decks)},
  author       = {Dorafshan, Sattar and Azari, Hoda},
  howpublished = {Mendeley Data},
  note         = {Version 1},
  year         = {2020},
  doi          = {10.17632/44rb96872r.1}
}

@article{young1995recursive,
  title   = {Recursive implementation of the {Gaussian} filter},
  author  = {Young, Ian T. and van Vliet, Lucas J.},
  journal = {Signal Processing},
  volume  = {44},
  number  = {2},
  pages   = {139--151},
  year    = {1995},
  doi     = {10.1016/0165-1684(95)00020-E}
}

@article{yeh2008wavelet,
  title   = {Application of the wavelet transform and the enhanced {Fourier} spectrum in the impact echo test},
  author  = {Yeh, Po-Liang and Liu, Pei-Ling},
  journal = {NDT \& E International},
  volume  = {41},
  number  = {5},
  pages   = {382--394},
  year    = {2008},
  doi     = {10.1016/j.ndteint.2008.01.002}
}

@article{zhang2012eemd,
  title   = {Ensemble empirical mode decomposition of impact-echo data for testing concrete structures},
  author  = {Zhang, Ying and Xie, Zhenhua},
  journal = {NDT \& E International},
  volume  = {51},
  pages   = {74--84},
  year    = {2012},
  doi     = {10.1016/j.ndteint.2012.05.008}
}

@misc{kim2024filtration,
  title        = {Filtration learning in exact multi-parameter persistent homology and classification of time-series data},
  author       = {Kim, Keunsu and Jung, Jae-Hun},
  howpublished = {arXiv preprint arXiv:2406.19587},
  year         = {2024},
  doi          = {10.48550/arXiv.2406.19587}
}

@article{lilly2012generalized,
  title   = {Generalized {Morse} wavelets as a superfamily of analytic wavelets},
  author  = {Lilly, Jonathan M. and Olhede, Sofia C.},
  journal = {IEEE Transactions on Signal Processing},
  volume  = {60},
  number  = {11},
  pages   = {6036--6041},
  year    = {2012},
  doi     = {10.1109/TSP.2012.2210890}
}

@article{xiang2002detection,
  title   = {Detection and classification of flaws in concrete structure using bispectra and neural networks},
  author  = {Xiang, Yang and Tso, Shiu Kit},
  journal = {NDT \& E International},
  volume  = {35},
  number  = {1},
  pages   = {19--27},
  year    = {2002},
  doi     = {10.1016/S0963-8695(01)00018-4}
}

@article{dorafshan2020evaluation,
  title   = {Evaluation of bridge decks with overlays using impact echo, a deep learning approach},
  author  = {Dorafshan, Sattar and Azari, Hoda},
  journal = {Automation in Construction},
  volume  = {113},
  pages   = {103133},
  year    = {2020},
  doi     = {10.1016/j.autcon.2020.103133}
}

\appendix

\clearpage
\onecolumn
\begingroup
\setcounter{table}{0}
\renewcommand{\thetable}{S\arabic{table}}
\providecommand{\theHtable}{\arabic{table}}
\renewcommand{\theHtable}{supp.\arabic{table}}
\setlength{\parindent}{0pt}
\setlength{\parskip}{6pt}
\setlength{\LTleft}{\fill}
\setlength{\LTright}{\fill}
\setlength{\tabcolsep}{7pt}
\renewcommand{\arraystretch}{1.10}
\raggedbottom

\begin{center}
{\Large\bfseries Supplementary Material\par}
\end{center}
As a supplementary analysis, we evaluated three-class classification on the experimental backfill-void dataset, treating measurement
regions A, B, and C as separate classes.
The results are reported alongside the binary classification results in Tables~\ref{tab:supp_fig5}--\ref{tab:supp_f1}.

Values are percentages and are reported as the mean \(\pm\) standard deviation across five fold-specific models evaluated on the same held-out
test set within each dataset.
All four methods use the same type of linear classifier with a Softmax output; the Fourier-derived representations use \(N=500\).
Precision, recall, and F1-score are macro-averaged over classes before aggregation across models.
Original signals have no artificially added noise.
For each noisy condition, the noise type and signal-to-noise ratio (SNR) are matched across training, validation, and testing.

\small
\begin{longtable}{@{}lrrrr@{}}
\caption{Condition-specific classification accuracy (\%). Values are means $\pm$ standard deviations across five fold-specific models evaluated on the same held-out test set.}
\label{tab:supp_fig5}\\
\midrule
Condition & FA-PI & FFT & Smoothed FFT & CWT \\
\midrule
\endfirsthead

\multicolumn{5}{c}{\tablename\ \thetable\ -- continued from previous page} \\
\midrule
Condition & FA-PI & FFT & Smoothed FFT & CWT \\
\midrule
\endhead

\midrule
\multicolumn{5}{r}{Continued on next page} \\
\endfoot

\midrule
\endlastfoot

\multicolumn{5}{l}{\textbf{Experimental backfill-void dataset: binary classification}} \\
\midrule
Original signals & 99.3 $\pm$ 0.1 & 99.3 $\pm$ 0.4 & 99.3 $\pm$ 0.4 & 99.7 $\pm$ 0.2 \\
Gaussian 20 dB & 99.0 $\pm$ 0.3 & 97.7 $\pm$ 0.3 & 97.7 $\pm$ 0.3 & 98.2 $\pm$ 0.9 \\
Gaussian 10 dB & 95.0 $\pm$ 0.3 & 93.9 $\pm$ 0.3 & 93.9 $\pm$ 0.3 & 95.8 $\pm$ 0.2 \\
Gaussian 5 dB & 90.2 $\pm$ 0.2 & 88.9 $\pm$ 0.7 & 88.9 $\pm$ 0.7 & 89.2 $\pm$ 0.5 \\
Gaussian 0 dB & 81.2 $\pm$ 0.6 & 80.5 $\pm$ 0.8 & 79.8 $\pm$ 0.7 & 80.2 $\pm$ 0.4 \\
Sinusoidal 20 dB & 98.2 $\pm$ 0.3 & 97.8 $\pm$ 0.4 & 97.8 $\pm$ 0.4 & 98.7 $\pm$ 0.4 \\
Sinusoidal 10 dB & 96.7 $\pm$ 0.2 & 92.4 $\pm$ 1.1 & 92.4 $\pm$ 1.1 & 94.5 $\pm$ 1.2 \\
Sinusoidal 5 dB & 96.5 $\pm$ 0.8 & 85.4 $\pm$ 1.3 & 83.9 $\pm$ 0.6 & 90.5 $\pm$ 1.8 \\
Sinusoidal 0 dB & 94.8 $\pm$ 0.3 & 75.9 $\pm$ 1.7 & 74.6 $\pm$ 0.9 & 85.1 $\pm$ 1.6 \\
Mixed 20 dB & 98.5 $\pm$ 0.3 & 97.7 $\pm$ 0.4 & 97.7 $\pm$ 0.4 & 98.6 $\pm$ 0.4 \\
Mixed 10 dB & 94.0 $\pm$ 0.5 & 92.1 $\pm$ 0.4 & 92.1 $\pm$ 0.4 & 94.3 $\pm$ 1.4 \\
Mixed 5 dB & 89.3 $\pm$ 0.9 & 85.2 $\pm$ 0.7 & 85.0 $\pm$ 0.6 & 87.2 $\pm$ 0.6 \\
Mixed 0 dB & 81.6 $\pm$ 0.9 & 74.7 $\pm$ 1.3 & 75.0 $\pm$ 0.8 & 77.7 $\pm$ 0.7 \\

\midrule
\multicolumn{5}{l}{\textbf{Public bridge-deck dataset: binary classification}} \\
\midrule
Original signals & 88.2 $\pm$ 1.6 & 83.3 $\pm$ 0.5 & 83.3 $\pm$ 0.5 & 89.3 $\pm$ 0.5 \\
Gaussian 20 dB & 87.8 $\pm$ 1.2 & 83.2 $\pm$ 0.4 & 83.2 $\pm$ 0.4 & 88.6 $\pm$ 0.8 \\
Gaussian 10 dB & 87.5 $\pm$ 1.1 & 83.3 $\pm$ 0.3 & 83.3 $\pm$ 0.3 & 88.1 $\pm$ 1.4 \\
Gaussian 5 dB & 86.6 $\pm$ 0.9 & 82.3 $\pm$ 0.6 & 82.5 $\pm$ 0.3 & 87.6 $\pm$ 1.0 \\
Gaussian 0 dB & 84.2 $\pm$ 0.7 & 81.2 $\pm$ 0.3 & 81.1 $\pm$ 0.4 & 86.0 $\pm$ 0.6 \\
Sinusoidal 20 dB & 87.9 $\pm$ 1.2 & 83.3 $\pm$ 0.6 & 83.3 $\pm$ 0.6 & 88.6 $\pm$ 0.6 \\
Sinusoidal 10 dB & 86.6 $\pm$ 0.7 & 82.1 $\pm$ 0.7 & 82.1 $\pm$ 0.7 & 87.5 $\pm$ 0.8 \\
Sinusoidal 5 dB & 85.3 $\pm$ 0.4 & 80.9 $\pm$ 0.5 & 80.9 $\pm$ 0.5 & 86.7 $\pm$ 1.2 \\
Sinusoidal 0 dB & 84.5 $\pm$ 1.2 & 78.8 $\pm$ 0.6 & 79.1 $\pm$ 0.3 & 85.6 $\pm$ 0.8 \\
Mixed 20 dB & 87.6 $\pm$ 0.8 & 83.1 $\pm$ 0.7 & 83.1 $\pm$ 0.7 & 88.8 $\pm$ 0.2 \\
Mixed 10 dB & 87.4 $\pm$ 1.0 & 82.3 $\pm$ 0.5 & 82.3 $\pm$ 0.5 & 87.5 $\pm$ 1.2 \\
Mixed 5 dB & 85.3 $\pm$ 0.8 & 81.4 $\pm$ 0.7 & 81.4 $\pm$ 0.7 & 87.1 $\pm$ 0.6 \\
Mixed 0 dB & 84.4 $\pm$ 1.1 & 80.7 $\pm$ 0.8 & 80.0 $\pm$ 0.5 & 85.6 $\pm$ 0.1 \\

\midrule
\multicolumn{5}{l}{\textbf{Experimental backfill-void dataset: three-class classification}} \\
\midrule
Original signals & 94.4 $\pm$ 0.4 & 90.7 $\pm$ 0.5 & 90.7 $\pm$ 0.5 & 96.9 $\pm$ 0.2 \\
Gaussian 20 dB & 92.5 $\pm$ 0.6 & 89.0 $\pm$ 0.4 & 89.0 $\pm$ 0.4 & 89.1 $\pm$ 0.7 \\
Gaussian 10 dB & 86.8 $\pm$ 0.5 & 83.1 $\pm$ 0.4 & 83.1 $\pm$ 0.4 & 82.0 $\pm$ 1.0 \\
Gaussian 5 dB & 78.4 $\pm$ 0.4 & 77.0 $\pm$ 0.6 & 76.8 $\pm$ 0.6 & 75.2 $\pm$ 0.7 \\
Gaussian 0 dB & 67.7 $\pm$ 1.2 & 66.7 $\pm$ 0.8 & 67.3 $\pm$ 0.8 & 66.0 $\pm$ 0.4 \\
Sinusoidal 20 dB & 91.0 $\pm$ 0.4 & 88.2 $\pm$ 0.2 & 88.2 $\pm$ 0.2 & 89.3 $\pm$ 1.3 \\
Sinusoidal 10 dB & 86.1 $\pm$ 0.5 & 81.1 $\pm$ 2.1 & 81.1 $\pm$ 2.1 & 81.5 $\pm$ 1.3 \\
Sinusoidal 5 dB & 84.4 $\pm$ 0.6 & 73.6 $\pm$ 1.5 & 73.6 $\pm$ 1.5 & 76.5 $\pm$ 0.8 \\
Sinusoidal 0 dB & 82.6 $\pm$ 0.6 & 64.8 $\pm$ 1.7 & 63.7 $\pm$ 1.5 & 70.4 $\pm$ 0.7 \\
Mixed 20 dB & 91.3 $\pm$ 0.5 & 88.4 $\pm$ 0.5 & 88.4 $\pm$ 0.5 & 87.6 $\pm$ 1.0 \\
Mixed 10 dB & 83.9 $\pm$ 0.8 & 80.9 $\pm$ 0.9 & 80.9 $\pm$ 0.8 & 79.5 $\pm$ 0.8 \\
Mixed 5 dB & 78.1 $\pm$ 0.7 & 73.6 $\pm$ 0.7 & 73.4 $\pm$ 0.5 & 72.3 $\pm$ 0.4 \\
Mixed 0 dB & 69.7 $\pm$ 0.3 & 62.8 $\pm$ 0.9 & 62.6 $\pm$ 0.6 & 62.6 $\pm$ 0.5 \\

\end{longtable}

\begin{longtable}{@{}lrrrr@{}}
\caption{Condition-specific macro-averaged precision (\%). Values are means $\pm$ standard deviations across five fold-specific models evaluated on the same held-out test set.}
\label{tab:supp_precision}\\
\midrule
Condition & FA-PI & FFT & Smoothed FFT & CWT \\
\midrule
\endfirsthead

\multicolumn{5}{c}{\tablename\ \thetable\ -- continued from previous page} \\
\midrule
Condition & FA-PI & FFT & Smoothed FFT & CWT \\
\midrule
\endhead

\midrule
\multicolumn{5}{r}{Continued on next page} \\
\endfoot

\midrule
\endlastfoot

\multicolumn{5}{l}{\textbf{Experimental backfill-void dataset: binary classification}} \\
\midrule
Original signals & 99.3 $\pm$ 0.1 & 99.4 $\pm$ 0.4 & 99.4 $\pm$ 0.4 & 99.7 $\pm$ 0.2 \\
Gaussian 20 dB & 99.0 $\pm$ 0.3 & 97.7 $\pm$ 0.3 & 97.7 $\pm$ 0.3 & 98.2 $\pm$ 1.0 \\
Gaussian 10 dB & 95.1 $\pm$ 0.2 & 94.0 $\pm$ 0.3 & 94.0 $\pm$ 0.3 & 95.9 $\pm$ 0.2 \\
Gaussian 5 dB & 90.2 $\pm$ 0.2 & 89.0 $\pm$ 0.7 & 89.0 $\pm$ 0.7 & 89.3 $\pm$ 0.5 \\
Gaussian 0 dB & 81.3 $\pm$ 0.5 & 80.6 $\pm$ 0.8 & 80.0 $\pm$ 0.9 & 80.3 $\pm$ 0.4 \\
Sinusoidal 20 dB & 98.2 $\pm$ 0.3 & 97.8 $\pm$ 0.4 & 97.8 $\pm$ 0.4 & 98.7 $\pm$ 0.3 \\
Sinusoidal 10 dB & 96.8 $\pm$ 0.2 & 92.5 $\pm$ 1.1 & 92.5 $\pm$ 1.1 & 94.5 $\pm$ 1.2 \\
Sinusoidal 5 dB & 96.5 $\pm$ 0.8 & 85.4 $\pm$ 1.2 & 84.0 $\pm$ 0.6 & 90.7 $\pm$ 1.5 \\
Sinusoidal 0 dB & 94.8 $\pm$ 0.3 & 75.9 $\pm$ 1.7 & 74.7 $\pm$ 1.0 & 85.3 $\pm$ 1.4 \\
Mixed 20 dB & 98.5 $\pm$ 0.3 & 97.7 $\pm$ 0.4 & 97.7 $\pm$ 0.4 & 98.6 $\pm$ 0.4 \\
Mixed 10 dB & 94.1 $\pm$ 0.5 & 92.1 $\pm$ 0.4 & 92.1 $\pm$ 0.4 & 94.3 $\pm$ 1.3 \\
Mixed 5 dB & 89.4 $\pm$ 0.8 & 85.2 $\pm$ 0.7 & 85.1 $\pm$ 0.5 & 87.4 $\pm$ 0.5 \\
Mixed 0 dB & 81.9 $\pm$ 0.8 & 75.0 $\pm$ 1.2 & 75.2 $\pm$ 0.7 & 78.1 $\pm$ 0.7 \\

\midrule
\multicolumn{5}{l}{\textbf{Public bridge-deck dataset: binary classification}} \\
\midrule
Original signals & 88.3 $\pm$ 1.8 & 85.1 $\pm$ 0.7 & 85.1 $\pm$ 0.7 & 90.5 $\pm$ 0.8 \\
Gaussian 20 dB & 88.1 $\pm$ 1.3 & 84.9 $\pm$ 0.7 & 84.9 $\pm$ 0.7 & 89.9 $\pm$ 0.9 \\
Gaussian 10 dB & 87.7 $\pm$ 0.7 & 85.4 $\pm$ 0.9 & 85.4 $\pm$ 0.9 & 90.3 $\pm$ 0.9 \\
Gaussian 5 dB & 86.2 $\pm$ 1.6 & 84.9 $\pm$ 0.9 & 85.0 $\pm$ 0.9 & 90.5 $\pm$ 0.6 \\
Gaussian 0 dB & 83.7 $\pm$ 0.2 & 83.4 $\pm$ 1.3 & 83.9 $\pm$ 0.6 & 89.5 $\pm$ 0.4 \\
Sinusoidal 20 dB & 87.7 $\pm$ 0.8 & 85.2 $\pm$ 1.0 & 85.2 $\pm$ 1.0 & 90.0 $\pm$ 0.7 \\
Sinusoidal 10 dB & 86.1 $\pm$ 1.0 & 83.3 $\pm$ 0.7 & 83.3 $\pm$ 0.7 & 89.6 $\pm$ 0.9 \\
Sinusoidal 5 dB & 85.0 $\pm$ 1.1 & 80.8 $\pm$ 0.9 & 80.8 $\pm$ 0.9 & 88.4 $\pm$ 1.3 \\
Sinusoidal 0 dB & 83.6 $\pm$ 3.0 & 76.1 $\pm$ 1.4 & 78.4 $\pm$ 1.3 & 87.4 $\pm$ 0.7 \\
Mixed 20 dB & 88.0 $\pm$ 0.6 & 84.9 $\pm$ 0.9 & 84.9 $\pm$ 0.9 & 90.3 $\pm$ 1.2 \\
Mixed 10 dB & 86.8 $\pm$ 1.5 & 83.9 $\pm$ 1.3 & 83.9 $\pm$ 1.3 & 90.5 $\pm$ 1.4 \\
Mixed 5 dB & 85.4 $\pm$ 1.7 & 82.2 $\pm$ 1.5 & 82.2 $\pm$ 1.5 & 90.2 $\pm$ 0.8 \\
Mixed 0 dB & 84.5 $\pm$ 1.7 & 81.9 $\pm$ 1.5 & 80.5 $\pm$ 1.5 & 88.1 $\pm$ 0.9 \\

\midrule
\multicolumn{5}{l}{\textbf{Experimental backfill-void dataset: three-class classification}} \\
\midrule
Original signals & 92.9 $\pm$ 0.5 & 88.1 $\pm$ 0.6 & 88.1 $\pm$ 0.6 & 96.2 $\pm$ 0.3 \\
Gaussian 20 dB & 90.5 $\pm$ 0.9 & 86.6 $\pm$ 0.6 & 86.6 $\pm$ 0.6 & 86.4 $\pm$ 1.2 \\
Gaussian 10 dB & 83.9 $\pm$ 0.8 & 79.5 $\pm$ 0.6 & 79.5 $\pm$ 0.6 & 77.7 $\pm$ 1.3 \\
Gaussian 5 dB & 75.2 $\pm$ 0.5 & 72.7 $\pm$ 0.6 & 72.6 $\pm$ 0.6 & 70.0 $\pm$ 0.9 \\
Gaussian 0 dB & 62.7 $\pm$ 1.9 & 61.8 $\pm$ 1.3 & 62.5 $\pm$ 1.4 & 59.0 $\pm$ 0.4 \\
Sinusoidal 20 dB & 88.6 $\pm$ 0.6 & 85.4 $\pm$ 0.7 & 85.4 $\pm$ 0.7 & 86.9 $\pm$ 1.8 \\
Sinusoidal 10 dB & 82.8 $\pm$ 0.4 & 77.4 $\pm$ 1.8 & 77.4 $\pm$ 1.8 & 78.1 $\pm$ 1.6 \\
Sinusoidal 5 dB & 80.7 $\pm$ 0.7 & 69.5 $\pm$ 1.9 & 69.5 $\pm$ 1.9 & 72.1 $\pm$ 1.2 \\
Sinusoidal 0 dB & 78.6 $\pm$ 1.1 & 59.6 $\pm$ 2.4 & 57.7 $\pm$ 2.5 & 64.7 $\pm$ 1.0 \\
Mixed 20 dB & 89.0 $\pm$ 0.5 & 86.1 $\pm$ 0.7 & 86.1 $\pm$ 0.7 & 84.5 $\pm$ 1.1 \\
Mixed 10 dB & 80.5 $\pm$ 1.2 & 77.6 $\pm$ 1.5 & 77.7 $\pm$ 1.2 & 74.9 $\pm$ 0.7 \\
Mixed 5 dB & 74.1 $\pm$ 1.7 & 69.3 $\pm$ 1.0 & 69.0 $\pm$ 0.7 & 67.0 $\pm$ 1.2 \\
Mixed 0 dB & 64.8 $\pm$ 1.1 & 57.4 $\pm$ 1.2 & 58.0 $\pm$ 2.1 & 55.0 $\pm$ 1.7 \\

\end{longtable}

\begin{longtable}{@{}lrrrr@{}}
\caption{Condition-specific macro-averaged recall (\%). Values are means $\pm$ standard deviations across five fold-specific models evaluated on the same held-out test set.}
\label{tab:supp_recall}\\
\midrule
Condition & FA-PI & FFT & Smoothed FFT & CWT \\
\midrule
\endfirsthead

\multicolumn{5}{c}{\tablename\ \thetable\ -- continued from previous page} \\
\midrule
Condition & FA-PI & FFT & Smoothed FFT & CWT \\
\midrule
\endhead

\midrule
\multicolumn{5}{r}{Continued on next page} \\
\endfoot

\midrule
\endlastfoot

\multicolumn{5}{l}{\textbf{Experimental backfill-void dataset: binary classification}} \\
\midrule
Original signals & 99.3 $\pm$ 0.1 & 99.3 $\pm$ 0.5 & 99.3 $\pm$ 0.5 & 99.7 $\pm$ 0.2 \\
Gaussian 20 dB & 98.9 $\pm$ 0.3 & 97.7 $\pm$ 0.3 & 97.7 $\pm$ 0.3 & 98.2 $\pm$ 0.9 \\
Gaussian 10 dB & 95.0 $\pm$ 0.3 & 93.8 $\pm$ 0.3 & 93.8 $\pm$ 0.3 & 95.8 $\pm$ 0.3 \\
Gaussian 5 dB & 90.2 $\pm$ 0.3 & 88.8 $\pm$ 0.7 & 88.8 $\pm$ 0.7 & 89.2 $\pm$ 0.5 \\
Gaussian 0 dB & 81.1 $\pm$ 0.6 & 80.4 $\pm$ 0.8 & 79.6 $\pm$ 0.7 & 80.1 $\pm$ 0.3 \\
Sinusoidal 20 dB & 98.2 $\pm$ 0.3 & 97.7 $\pm$ 0.4 & 97.7 $\pm$ 0.4 & 98.7 $\pm$ 0.4 \\
Sinusoidal 10 dB & 96.7 $\pm$ 0.2 & 92.4 $\pm$ 1.1 & 92.4 $\pm$ 1.1 & 94.5 $\pm$ 1.1 \\
Sinusoidal 5 dB & 96.4 $\pm$ 0.8 & 85.3 $\pm$ 1.3 & 83.8 $\pm$ 0.6 & 90.5 $\pm$ 1.9 \\
Sinusoidal 0 dB & 94.8 $\pm$ 0.3 & 75.8 $\pm$ 1.7 & 74.5 $\pm$ 0.9 & 85.0 $\pm$ 1.7 \\
Mixed 20 dB & 98.5 $\pm$ 0.3 & 97.7 $\pm$ 0.4 & 97.7 $\pm$ 0.4 & 98.6 $\pm$ 0.4 \\
Mixed 10 dB & 94.0 $\pm$ 0.5 & 92.0 $\pm$ 0.4 & 92.0 $\pm$ 0.4 & 94.3 $\pm$ 1.3 \\
Mixed 5 dB & 89.2 $\pm$ 0.9 & 85.1 $\pm$ 0.7 & 85.0 $\pm$ 0.6 & 87.1 $\pm$ 0.6 \\
Mixed 0 dB & 81.5 $\pm$ 1.0 & 74.5 $\pm$ 1.3 & 74.8 $\pm$ 0.9 & 77.5 $\pm$ 0.7 \\

\midrule
\multicolumn{5}{l}{\textbf{Public bridge-deck dataset: binary classification}} \\
\midrule
Original signals & 79.5 $\pm$ 3.1 & 69.2 $\pm$ 1.3 & 69.2 $\pm$ 1.3 & 80.6 $\pm$ 0.8 \\
Gaussian 20 dB & 78.6 $\pm$ 2.4 & 69.0 $\pm$ 1.2 & 69.0 $\pm$ 1.2 & 79.4 $\pm$ 1.6 \\
Gaussian 10 dB & 78.1 $\pm$ 2.5 & 69.0 $\pm$ 0.8 & 69.0 $\pm$ 0.8 & 78.0 $\pm$ 2.9 \\
Gaussian 5 dB & 77.0 $\pm$ 2.4 & 66.9 $\pm$ 1.2 & 67.2 $\pm$ 0.9 & 76.7 $\pm$ 2.2 \\
Gaussian 0 dB & 72.1 $\pm$ 2.0 & 64.8 $\pm$ 1.0 & 64.3 $\pm$ 1.1 & 73.3 $\pm$ 1.3 \\
Sinusoidal 20 dB & 79.1 $\pm$ 2.9 & 69.2 $\pm$ 1.4 & 69.2 $\pm$ 1.4 & 79.3 $\pm$ 1.2 \\
Sinusoidal 10 dB & 76.8 $\pm$ 1.4 & 67.1 $\pm$ 1.5 & 67.1 $\pm$ 1.5 & 77.0 $\pm$ 1.8 \\
Sinusoidal 5 dB & 74.2 $\pm$ 1.1 & 65.0 $\pm$ 1.2 & 65.0 $\pm$ 1.2 & 75.6 $\pm$ 2.4 \\
Sinusoidal 0 dB & 73.3 $\pm$ 1.1 & 61.6 $\pm$ 1.5 & 61.5 $\pm$ 0.6 & 73.6 $\pm$ 1.9 \\
Mixed 20 dB & 78.3 $\pm$ 1.9 & 68.9 $\pm$ 1.6 & 68.9 $\pm$ 1.6 & 79.6 $\pm$ 0.9 \\
Mixed 10 dB & 78.7 $\pm$ 2.0 & 67.3 $\pm$ 1.3 & 67.2 $\pm$ 1.3 & 76.4 $\pm$ 2.3 \\
Mixed 5 dB & 74.1 $\pm$ 1.9 & 65.8 $\pm$ 1.3 & 65.8 $\pm$ 1.3 & 75.6 $\pm$ 1.1 \\
Mixed 0 dB & 72.1 $\pm$ 2.3 & 64.1 $\pm$ 1.8 & 62.8 $\pm$ 1.2 & 73.1 $\pm$ 0.5 \\

\midrule
\multicolumn{5}{l}{\textbf{Experimental backfill-void dataset: three-class classification}} \\
\midrule
Original signals & 92.3 $\pm$ 0.5 & 86.8 $\pm$ 0.8 & 86.8 $\pm$ 0.8 & 95.8 $\pm$ 0.4 \\
Gaussian 20 dB & 90.2 $\pm$ 0.8 & 84.7 $\pm$ 0.5 & 84.7 $\pm$ 0.5 & 85.6 $\pm$ 0.8 \\
Gaussian 10 dB & 83.4 $\pm$ 0.4 & 77.4 $\pm$ 0.8 & 77.4 $\pm$ 0.8 & 75.7 $\pm$ 1.4 \\
Gaussian 5 dB & 72.7 $\pm$ 1.2 & 70.1 $\pm$ 1.2 & 69.7 $\pm$ 1.2 & 67.1 $\pm$ 1.5 \\
Gaussian 0 dB & 59.8 $\pm$ 1.9 & 57.9 $\pm$ 1.2 & 58.8 $\pm$ 1.7 & 56.3 $\pm$ 1.1 \\
Sinusoidal 20 dB & 88.0 $\pm$ 0.6 & 83.5 $\pm$ 0.4 & 83.5 $\pm$ 0.4 & 85.9 $\pm$ 2.0 \\
Sinusoidal 10 dB & 81.0 $\pm$ 1.0 & 75.3 $\pm$ 2.9 & 75.3 $\pm$ 2.9 & 75.3 $\pm$ 1.7 \\
Sinusoidal 5 dB & 78.7 $\pm$ 1.0 & 66.5 $\pm$ 2.0 & 66.5 $\pm$ 2.0 & 68.6 $\pm$ 1.2 \\
Sinusoidal 0 dB & 76.5 $\pm$ 0.4 & 55.9 $\pm$ 2.1 & 54.5 $\pm$ 1.8 & 61.9 $\pm$ 1.3 \\
Mixed 20 dB & 88.5 $\pm$ 0.9 & 83.7 $\pm$ 0.7 & 83.7 $\pm$ 0.7 & 82.8 $\pm$ 1.8 \\
Mixed 10 dB & 78.4 $\pm$ 0.9 & 74.8 $\pm$ 1.2 & 74.8 $\pm$ 1.0 & 71.8 $\pm$ 1.7 \\
Mixed 5 dB & 71.8 $\pm$ 1.3 & 66.2 $\pm$ 1.1 & 65.6 $\pm$ 0.7 & 63.0 $\pm$ 1.1 \\
Mixed 0 dB & 62.3 $\pm$ 0.7 & 53.6 $\pm$ 1.3 & 52.4 $\pm$ 0.8 & 52.5 $\pm$ 1.2 \\

\end{longtable}

\begin{longtable}{@{}lrrrr@{}}
\caption{Condition-specific macro-averaged F1-score (\%). Values are means $\pm$ standard deviations across five fold-specific models evaluated on the same held-out test set.}
\label{tab:supp_f1}\\
\midrule
Condition & FA-PI & FFT & Smoothed FFT & CWT \\
\midrule
\endfirsthead

\multicolumn{5}{c}{\tablename\ \thetable\ -- continued from previous page} \\
\midrule
Condition & FA-PI & FFT & Smoothed FFT & CWT \\
\midrule
\endhead

\midrule
\multicolumn{5}{r}{Continued on next page} \\
\endfoot

\midrule
\endlastfoot

\multicolumn{5}{l}{\textbf{Experimental backfill-void dataset: binary classification}} \\
\midrule
Original signals & 99.3 $\pm$ 0.1 & 99.3 $\pm$ 0.4 & 99.3 $\pm$ 0.4 & 99.7 $\pm$ 0.2 \\
Gaussian 20 dB & 99.0 $\pm$ 0.3 & 97.7 $\pm$ 0.3 & 97.7 $\pm$ 0.3 & 98.2 $\pm$ 0.9 \\
Gaussian 10 dB & 95.0 $\pm$ 0.3 & 93.9 $\pm$ 0.3 & 93.9 $\pm$ 0.3 & 95.8 $\pm$ 0.2 \\
Gaussian 5 dB & 90.2 $\pm$ 0.3 & 88.8 $\pm$ 0.7 & 88.8 $\pm$ 0.7 & 89.2 $\pm$ 0.5 \\
Gaussian 0 dB & 81.1 $\pm$ 0.6 & 80.4 $\pm$ 0.8 & 79.7 $\pm$ 0.7 & 80.2 $\pm$ 0.3 \\
Sinusoidal 20 dB & 98.2 $\pm$ 0.3 & 97.8 $\pm$ 0.4 & 97.8 $\pm$ 0.4 & 98.7 $\pm$ 0.4 \\
Sinusoidal 10 dB & 96.7 $\pm$ 0.2 & 92.4 $\pm$ 1.1 & 92.4 $\pm$ 1.1 & 94.5 $\pm$ 1.2 \\
Sinusoidal 5 dB & 96.4 $\pm$ 0.8 & 85.3 $\pm$ 1.3 & 83.8 $\pm$ 0.6 & 90.5 $\pm$ 1.9 \\
Sinusoidal 0 dB & 94.8 $\pm$ 0.3 & 75.8 $\pm$ 1.7 & 74.5 $\pm$ 0.9 & 85.0 $\pm$ 1.7 \\
Mixed 20 dB & 98.5 $\pm$ 0.3 & 97.7 $\pm$ 0.4 & 97.7 $\pm$ 0.4 & 98.6 $\pm$ 0.4 \\
Mixed 10 dB & 94.0 $\pm$ 0.5 & 92.1 $\pm$ 0.4 & 92.1 $\pm$ 0.4 & 94.3 $\pm$ 1.4 \\
Mixed 5 dB & 89.2 $\pm$ 0.9 & 85.1 $\pm$ 0.7 & 85.0 $\pm$ 0.6 & 87.1 $\pm$ 0.6 \\
Mixed 0 dB & 81.5 $\pm$ 1.0 & 74.5 $\pm$ 1.3 & 74.8 $\pm$ 0.9 & 77.6 $\pm$ 0.8 \\

\midrule
\multicolumn{5}{l}{\textbf{Public bridge-deck dataset: binary classification}} \\
\midrule
Original signals & 82.5 $\pm$ 2.8 & 72.5 $\pm$ 1.3 & 72.5 $\pm$ 1.3 & 84.0 $\pm$ 0.8 \\
Gaussian 20 dB & 81.8 $\pm$ 2.2 & 72.2 $\pm$ 1.3 & 72.2 $\pm$ 1.3 & 82.9 $\pm$ 1.4 \\
Gaussian 10 dB & 81.2 $\pm$ 2.2 & 72.3 $\pm$ 0.9 & 72.3 $\pm$ 0.9 & 81.7 $\pm$ 2.7 \\
Gaussian 5 dB & 79.9 $\pm$ 2.0 & 69.9 $\pm$ 1.4 & 70.2 $\pm$ 1.0 & 80.6 $\pm$ 2.1 \\
Gaussian 0 dB & 75.2 $\pm$ 1.9 & 67.2 $\pm$ 1.3 & 66.7 $\pm$ 1.4 & 77.3 $\pm$ 1.3 \\
Sinusoidal 20 dB & 82.0 $\pm$ 2.4 & 72.5 $\pm$ 1.5 & 72.5 $\pm$ 1.5 & 82.8 $\pm$ 1.1 \\
Sinusoidal 10 dB & 79.9 $\pm$ 1.3 & 70.0 $\pm$ 1.8 & 70.0 $\pm$ 1.8 & 80.7 $\pm$ 1.6 \\
Sinusoidal 5 dB & 77.4 $\pm$ 0.9 & 67.4 $\pm$ 1.4 & 67.4 $\pm$ 1.4 & 79.2 $\pm$ 2.4 \\
Sinusoidal 0 dB & 76.3 $\pm$ 1.3 & 63.0 $\pm$ 1.9 & 62.9 $\pm$ 0.9 & 77.2 $\pm$ 1.7 \\
Mixed 20 dB & 81.5 $\pm$ 1.7 & 72.0 $\pm$ 1.8 & 72.0 $\pm$ 1.8 & 83.1 $\pm$ 0.5 \\
Mixed 10 dB & 81.5 $\pm$ 1.7 & 70.2 $\pm$ 1.5 & 70.1 $\pm$ 1.5 & 80.3 $\pm$ 2.3 \\
Mixed 5 dB & 77.3 $\pm$ 1.7 & 68.4 $\pm$ 1.6 & 68.4 $\pm$ 1.6 & 79.6 $\pm$ 1.1 \\
Mixed 0 dB & 75.3 $\pm$ 2.2 & 66.3 $\pm$ 2.2 & 64.7 $\pm$ 1.5 & 76.9 $\pm$ 0.4 \\

\midrule
\multicolumn{5}{l}{\textbf{Experimental backfill-void dataset: three-class classification}} \\
\midrule
Original signals & 92.6 $\pm$ 0.5 & 87.3 $\pm$ 0.7 & 87.3 $\pm$ 0.7 & 95.9 $\pm$ 0.3 \\
Gaussian 20 dB & 90.3 $\pm$ 0.8 & 85.3 $\pm$ 0.5 & 85.3 $\pm$ 0.5 & 85.7 $\pm$ 0.7 \\
Gaussian 10 dB & 83.6 $\pm$ 0.6 & 77.8 $\pm$ 0.8 & 77.8 $\pm$ 0.8 & 76.1 $\pm$ 1.4 \\
Gaussian 5 dB & 73.2 $\pm$ 1.3 & 70.3 $\pm$ 1.3 & 69.9 $\pm$ 1.3 & 67.2 $\pm$ 1.8 \\
Gaussian 0 dB & 59.7 $\pm$ 2.0 & 57.5 $\pm$ 1.5 & 58.3 $\pm$ 2.1 & 54.6 $\pm$ 1.6 \\
Sinusoidal 20 dB & 88.2 $\pm$ 0.6 & 84.1 $\pm$ 0.4 & 84.1 $\pm$ 0.4 & 86.2 $\pm$ 1.9 \\
Sinusoidal 10 dB & 81.4 $\pm$ 1.0 & 75.6 $\pm$ 3.1 & 75.6 $\pm$ 3.1 & 75.5 $\pm$ 2.1 \\
Sinusoidal 5 dB & 79.2 $\pm$ 1.0 & 66.6 $\pm$ 2.2 & 66.6 $\pm$ 2.2 & 68.4 $\pm$ 1.5 \\
Sinusoidal 0 dB & 77.0 $\pm$ 0.5 & 55.2 $\pm$ 2.4 & 53.3 $\pm$ 2.2 & 61.2 $\pm$ 1.5 \\
Mixed 20 dB & 88.7 $\pm$ 0.8 & 84.4 $\pm$ 0.7 & 84.4 $\pm$ 0.7 & 83.3 $\pm$ 1.7 \\
Mixed 10 dB & 78.9 $\pm$ 1.0 & 75.4 $\pm$ 1.3 & 75.5 $\pm$ 1.1 & 71.7 $\pm$ 2.3 \\
Mixed 5 dB & 72.1 $\pm$ 1.5 & 66.5 $\pm$ 1.2 & 65.7 $\pm$ 0.9 & 61.5 $\pm$ 2.2 \\
Mixed 0 dB & 62.1 $\pm$ 0.8 & 52.7 $\pm$ 1.4 & 50.6 $\pm$ 1.1 & 50.5 $\pm$ 2.2 \\

\end{longtable}

\clearpage
\endgroup

\end{document}